\documentclass[12pt,a4paper]{article}
\usepackage{amsmath,amssymb}
\usepackage[dvipdfmx]{graphicx}
\usepackage[dvips]{color}
\begin{document}
%\draft
\title{
A Diagrammer's Note on 
Superconducting Fluctuation Transport 
for Beginners: \\
I. Conductivity and Thermopower 
}

\author{
O. Narikiyo
\footnote{
Department of Physics, 
Kyushu University, 
Fukuoka 819-0395, 
Japan}
}

\date{
(Sep. 3, 2026)
}

\maketitle
%----------------------------------------------------------------
\begin{abstract}
A diagrammatic approach 
based on thermal Green function 
to superconducting fluctuation transport 
is reviewed keeping consistency with Ginzburg-Landau theory. 
The correct expression of the heat current vertex for Cooper pairs is clarified via Jonson-Mahan formula and Ward identities. 
\end{abstract}
%----------------------------------------------------------------
\vskip 20pt

%%%%%%%%
\section{Introduction}
%%%%%%%%

In this Note, I, {\underline {a diagrammer}}, try to 
supply you, {\underline {beginners}}, 
with some detailed calculations 
usually skipped in the references written by professionals.\footnote{
I shall remark on some mistakes made by professionals in several footnotes. 
Such notes might be helpful for you to avoid unnecessary struggle 
with the mistakes. } 
I hope that you can save time and efforts by this note 
and reserve enough energy to go farther. 

This note is intended to be a primer\footnote{
In the discussion of superconducting fluctuation transport ($T > T_c$) 
I only consider the Aslamazov-Larkin (AL) contribution 
and leave the consideration of the Maki-Thompson (MT) and 
the density-of-states (DOS) contributions to the textbook \cite{old,new}. 
The AL transport is equivalent to 
the time-dependent Ginzburg-Landau (GL) transport. } 
to the textbook: 
{\it Theory of Fluctuations in Superconductors} \cite{old,new}. 
It is the first of the series consisting of three notes. 
This first note mainly discusses the transport 
in the absence of magnetic field. 
The second and the third mainly discuss the transport under magnetic field. 

I expect that you have some knowledge 
of the Feynman-diagram technique at finite temperature 
and the linear response calculation 
via the thermal Green function. 
I will skip detailed explanations of these in the following. 

We use the unit of $\hbar = k_B = c = 1$. 

%%%%%%%%
\section{Hamiltonian}
%%%%%%%%

In grand canonical formalism we use $K$ 
\begin{equation}
K = H - \mu N, 
\label{K} 
\end{equation}
instead of the Hamiltonian $H$ 
where $N$ and $\mu$ are the number and chemical potential of electrons. 
We decompose $K$ as 
\begin{equation}
K = K_0 + V, 
\end{equation}
where 
\begin{equation}
K_0 = \sum_{\bf p} \xi_{\bf p} \big(
a_{\bf p}^\dag a_{\bf p} + b_{\bf p}^\dag b_{\bf p} 
\big), 
\end{equation}
is the kinetic energy 
measured from the chemical potential $\mu$ with 
\begin{equation}
\xi_{\bf p} = {{\bf p}^2 \over 2m} - \mu, 
\label{xi(p)} 
\end{equation}
and 
\begin{equation}
V = - g \sum_{\bf q} P_{\bf q}^\dag P_{\bf q}, 
\label{local-g} 
\end{equation}
is the local attractive interaction with $g$ being a positive constant. 
Here $a_{\bf p}^\dag$ represents the creation operator 
of an up-spin electron with momentum ${\bf p}$, 
$b_{\bf p}$ represents the annihilation operator of a down-spin electron 
and so on. 
The creation and annihilation operators of Cooper Pairs are 
\begin{equation}
P_{\bf q}^\dag = \sum_{\bf p} 
a_{ {\bf p} + {{\bf q}\over 2} }^\dag b_{ -{\bf p} + {{\bf q}\over 2} }^\dag, 
\ \ \ \ \ \ \ \ \ \ 
P_{\bf q} = \sum_{\bf p} 
b_{ -{\bf p} + {{\bf q}\over 2} } a_{ {\bf p} + {{\bf q}\over 2} }. 
\end{equation}

The (imaginary) time dependence of the operator $A$ is given as 
\begin{equation}
A(\tau) = e^{K\tau} A e^{-K\tau}, 
\end{equation}
or 
\begin{equation}
{\partial \over \partial \tau}A(\tau) = \big[ K, A(\tau) \big]. 
\label{Eq-of-motion} 
\end{equation}

%%%%%%%%
\section{Electron Propagator}
%%%%%%%%

The propagator for electrons $G({\bf p}, \tau)$ is defined as 
\begin{equation}
G({\bf p}, \tau) = - 
\big\langle T_\tau \big\{ a_{\bf p}(\tau) 
                          a_{\bf p}^\dag \big\} \big\rangle, 
\end{equation}
where $\big\langle X \big\rangle$ 
is the expectation value of $X$ 
using the density matrix defined by $K$. 
In the following, throughout the series of three notes, the Zeeman splitting is neglected so that 
\begin{equation}
- 
\big\langle T_\tau \big\{ b_{\bf p}(\tau) 
                          b_{\bf p}^\dag \big\} \big\rangle 
= G({\bf p}, \tau). 
\end{equation}

The operation of the (imaginary) time ordering $T_\tau$ 
is explicitly expressed as 
\begin{equation}
G({\bf p}, \tau) = 
- \theta( \tau) \big\langle a_{\bf p}(\tau) a_{\bf p}^\dag \big\rangle 
+ \theta(-\tau) \big\langle a_{\bf p}^\dag a_{\bf p}(\tau) \big\rangle, 
\end{equation}
for electrons 
where 
$\theta(x)=1$ for $x>0$ and $\theta(x)=0$ for $x<0$. 
Since the time evolution of the free electron is given as\footnote{
$a_{\bf p}^\dag(\tau) = e^{\xi_{\bf p}\tau} a_{\bf p}^\dag$.} 
\begin{equation}
a_{\bf p}(\tau) = e^{-\xi_{\bf p}\tau} a_{\bf p}, 
\end{equation}
the propagator for free electrons $G_0({\bf p}, \tau)$ becomes\footnote{
$ a_{\bf p} a_{\bf p}^\dag = 1 - a_{\bf p}^\dag a_{\bf p} $. 
For free electrons 
$ \big\langle a_{\bf p}^\dag a_{\bf p} \big\rangle_0 = f(\xi_{\bf p}) $ 
where $\big\langle X \big\rangle_0$ 
is the expectation value of $X$ using the density matrix defined by $K_0$. } 
\begin{equation}
G_0({\bf p}, \tau) = e^{-\xi_{\bf p} \tau} \big\{ 
- \theta( \tau) \big[ 1 - f(\xi_{\bf p}) \big] 
+ \theta(-\tau) f(\xi_{\bf p}) \big\}, 
\end{equation}
where $f(x)$ is the Fermi distribution function 
\begin{equation}
f(x) = { 1 \over e^{\beta x} + 1 }. \label{FD}
\end{equation}
The full propagator $G({\bf p}, \tau)$ 
is expressed as the superposition of this-type of function$:$ 
\begin{equation}
G({\bf p}, \tau) = \int_{-\infty}^\infty d \epsilon \rho(\epsilon) 
e^{-\epsilon \tau} \big\{ 
- \theta( \tau) \big[ 1 - f(\epsilon) \big] 
+ \theta(-\tau) f(\epsilon) \big\}. 
\end{equation}
Here $\rho(\epsilon)$ is the spectral function. 

The Fourier transform of the propagator is defined as 
\begin{equation}
G({\bf p}, \tau) = {1 \over \beta} \sum_{n=-\infty}^\infty 
G({\bf p}, i\varepsilon_n) e^{-i\varepsilon_n \tau}, 
\end{equation}
\begin{equation}
G({\bf p}, i\varepsilon_n) = \int_0^\beta d \tau 
G({\bf p}, \tau) e^{i\varepsilon_n \tau}, 
\end{equation}
where $\varepsilon_n = (2n+1)\pi T$ 
with $n$ being integer. 
Here $T$ is the temperature and 
$\beta$ is the inverse temperature $\beta = 1/T$. 

The free propagator is transformed as 
\begin{equation}
G_0({\bf p}, i\varepsilon_n) = - \int_0^\beta d \tau 
\big\langle a_{\bf p}(\tau) a_{\bf p}^\dag \big\rangle 
e^{i\varepsilon_n \tau}
= - \big[ 1 - f(\xi_{\bf p}) \big] 
\int_0^\beta d \tau e^{(i\varepsilon_n - \xi_{\bf p})\tau}. 
\end{equation}
Using $e^{i\varepsilon_n \beta} = -1$, 
we obtain 
\begin{equation}
G_0({\bf p}, i\varepsilon_n) = 
{1 \over i\varepsilon_n - \xi_{\bf p}}. 
\end{equation}
The full propagator $G({\bf p}, i\varepsilon_n)$ 
is expressed as the superposition of this-type of function$:$ 
\begin{equation}
G({\bf p}, i\varepsilon_n) = \int_{-\infty}^\infty d \epsilon 
{\rho(\epsilon) \over i\varepsilon_n - \epsilon}. 
\label{sr-G} 
\end{equation}

The retarded and advanced propagators, 
$G^R({\bf p}, \varepsilon)$ and $G^A({\bf p}, \varepsilon)$, 
are obtained from the thermal propagator $G({\bf p}, i\varepsilon_n)$ as 
\begin{equation}
G^R({\bf p}, \varepsilon) = G({\bf p}, \varepsilon + i \delta), 
\ \ \ \ \ \ \ \ \ \ 
G^A({\bf p}, \varepsilon) = G({\bf p}, \varepsilon - i \delta), 
\end{equation}
where $\delta = +0$. 
The spectral function $\rho(\varepsilon)$ is related\footnote{
We have employed the identity 
\begin{equation}
{1 \over x + i\delta} = {{\rm P} \over x} - i \pi \delta(x), 
\nonumber 
\end{equation}
where ${\rm P}$ denotes that the principal value should be taken 
when it is integrated. } 
to the imaginary part of the retarded propagator as 
\begin{equation}
{\rm Im}G^R({\bf p}, \varepsilon) = - \pi \rho(\varepsilon). 
\end{equation}

The full and free propagators are related by the Dyson equation 
\begin{equation}
G({\bf p}, i\varepsilon_n)^{-1} = 
G_0({\bf p}, i\varepsilon_n)^{-1} - \Sigma({\bf p}, i\varepsilon_n), 
\end{equation}
where $\Sigma({\bf p}, i\varepsilon_n)$ 
is the self-energy. 

%%%%%%%%
\section{Cooper-Pair Propagator}
%%%%%%%%

%%%%%%%%%%%%%%%%%%%%%%%%%%%%%%%%%%%%%%%%%%
%\vskip 12mm
\begin{figure}[htbp]
\begin{center}
%\hskip -0.6cm
\includegraphics[width=16.0cm]{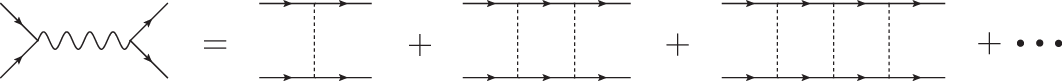}
\vskip 4mm
\caption{Effective interaction for Cooper pairs: 
The wavy and broken lines represent $-L$ and $-(-g)$. 
The solid line with arrow represents $G_0$. 
All the Feynman diagrams in this note are drawn by JaxoDraw. }
\label{fig:Fig-L}
\end{center}
\end{figure}
%%%%%%%%%%%%%%%%%%%%%%%%%%%%%%%%%%%%%%%%%%

The effective interaction for Cooper pairs 
in the ladder approximation $L({\bf q},i\omega_m)$ is given by 
\begin{align}
- L({\bf q},i\omega_m) & = g 
+ g \Pi({\bf q},i\omega_m) g 
+ g \Pi({\bf q},i\omega_m) g \Pi({\bf q},i\omega_m) g 
+ \cdot \cdot \cdot 
\nonumber \\ 
& = {g \over 1 - g \Pi({\bf q},i\omega_m)}, 
\label{ladder-L} 
\end{align}
where\footnote{
Since the particle-particle pair propagator $\Pi({\bf q},i\omega_m)$ 
depends on the shape of the region of ${\bf p}$-summation, 
we have chosen the symmetric configuration in ${\bf p}$-space 
as (\ref{Pi(q)}). 
On the other hand, 
the particle-hole pair propagator is free from the choice. 
See Fukuyama, Hasegawa and Narikiyo: J. Phys. Soc. Japan {\bf 60}, 2013 (1991) 
for these discussions. 
The imaginary part of $P^R(q,\omega)$ proportional to $q^2$ derived in 
Sergeev, Reizer and Mitin: Phys. Rev. B {\bf 66}, 104504 (2002) 
is absent in the symmetric choice. } 
\begin{equation}
\Pi({\bf q},i\omega_m) = T \sum_n \sum_{\bf p} 
G_0({\bf p} + {{\bf q}\over 2}, i\varepsilon_n) 
G_0(-{\bf p} + {{\bf q}\over 2}, i\omega_m - i\varepsilon_n), 
\label{Pi(q)} 
\end{equation}
and $\omega_m = 2m \pi T$ with $m$ being integer. 

The propagator for Cooper pairs $D({\bf q}, \tau)$ is defined as 
\begin{equation}
D({\bf q}, \tau) = - 
\big\langle T_\tau \big\{ P_{\bf q}(\tau) 
                          P_{\bf q}^\dag \big\} \big\rangle, 
\end{equation}
and its Fourier transform in the ladder approximation 
$D({\bf q},i\omega_m)$ is given by 
\begin{align}
- D({\bf q},i\omega_m) & = \Pi({\bf q},i\omega_m) 
+ \Pi({\bf q},i\omega_m) g \Pi({\bf q},i\omega_m) 
+ \cdot \cdot \cdot 
\nonumber \\ 
& = {\Pi({\bf q},i\omega_m) \over 1 - g \Pi({\bf q},i\omega_m)}. 
\end{align}

The relation\footnote{
In the forthcoming discussion employing the Ward identity 
the inverse of the propagator $D({\bf q},i\omega_m)^{-1}$ 
plays the central role. 
In the limit of $T \rightarrow T_c$, 
${\bf q} \rightarrow 0$ and $\omega_m \rightarrow 0$, 
\begin{equation}
D({\bf q}, i\omega_m)^{-1} 
= L({\bf q},i\omega_m)^{-1} \cdot {g \over \Pi({\bf q},i\omega_m)} 
\fallingdotseq 
  L({\bf q},i\omega_m)^{-1} \cdot {g \over \Pi(0,0)} 
\fallingdotseq 
L({\bf q},i\omega_m)^{-1} \cdot g^2, 
\nonumber
\end{equation}
where 
$ 1 - g \Pi(0,0) \sim O(T-T_c) $
and $ L({\bf q},i\omega_m)^{-1} \sim O(T-T_c, {\bf q}^2, \omega_m)$ as 
\begin{equation}
L({\bf q},i\omega_m)^{-1} = 
- \big[ g^{-1} - \Pi(0,0) \big] 
- \big[ \Pi(0,0) - \Pi({\bf q},i\omega_m) \big]. 
\nonumber
\end{equation}
Thus in this limit 
\begin{equation}
D_\Delta({\bf q}, i\omega_m)^{-1} 
\fallingdotseq 
L({\bf q},i\omega_m)^{-1}. 
\nonumber
\end{equation}
} 
between $D({\bf q},i\omega_m)$ and $L({\bf q},i\omega_m)$ is 
\begin{equation}
L({\bf q}, i\omega_m) = - g + g^2 D({\bf q},i\omega_m). 
\end{equation}
In terms of the propagator of the order-parameter 
$D_\Delta({\bf q},i\omega_m)$ 
this relation is rewritten as 
\begin{equation}
L({\bf q}, i\omega_m) = - g + D_\Delta({\bf q},i\omega_m), 
\end{equation}
where the propagator is introduced as 
\begin{equation}
D_\Delta({\bf q}, \tau) = - 
\big\langle T_\tau \big\{ \Delta_{\bf q}(\tau) 
                          \Delta_{\bf q}^\dag \big\} \big\rangle, 
\end{equation}
with 
\begin{equation}
\Delta_{\bf q}^\dag = g \cdot P_{\bf q}^\dag, 
\ \ \ \ \ \ \ \ \ \ 
\Delta_{\bf q} = g \cdot P_{\bf q}. 
\end{equation}

The retarded and advanced propagators, 
$D^R({\bf q}, \omega)$ and $D^R({\bf q}, \omega)$, 
are obtained from the thermal propagator $D({\bf q},i\omega_m)$ as 
\begin{equation}
D^R({\bf q},\omega) = D({\bf q}, \omega + i \delta), 
\ \ \ \ \ \ \ \ \ \ 
D^A({\bf q},\omega) = D({\bf q}, \omega - i \delta). 
\end{equation}

In the ladder approximation, 
the full and free propagators, 
$D({\bf q},i\omega_m)$ and $D_0({\bf q},i\omega_m)$, 
are related by the Dyson equation as 
\begin{equation}
D({\bf q},i\omega_m)^{-1} = 
D_0({\bf q},i\omega_m)^{-1} - \Sigma({\bf q},i\omega_m), 
\end{equation}
where 
\begin{equation}
D_0({\bf q},i\omega_m) = - \Pi({\bf q},i\omega_m), 
\ \ \ \ \ \ \ \ \ \ 
\Sigma({\bf q},i\omega_m) = - g. 
\end{equation}
In the limits of low energy and long wavelength 
the retarded effective interaction 
$ L^R({\bf q},\omega) = L({\bf q}, \omega + i \delta) $ 
obtained from (\ref{ladder-L}) is\footnote{
See \S 6.2 in \cite{new} for the microscopic calculation 
in the ladder approximation. 
In \S 6.4 of \cite{new} the renormalization due to impurity scattering 
is also discussed. } 
\begin{equation}
L^R({\bf q},\omega) = - {1 \over N(0)} 
{1 \over \epsilon + \xi_0^2 {\bf q}^2 - i \omega \tau_0}, 
\label{L-form} 
\end{equation}
where 
\begin{equation}
\epsilon \equiv \ln{T \over T_c} \fallingdotseq {T - T_c \over T_c}, 
\end{equation}
and $N(0)$ is the density of states per spin at the chemical potential. 
This effective interaction has 
Ornstein-Zernike form in space-direction and 
Debye form in time direction. 
The functional form of (\ref{L-form}) holds 
beyond the ladder approximation 
so that we regard $\epsilon$, $\xi_0$ and $\tau_0$ 
as phenomenological parameters in the following. 

%%%%%%%%
\section{Boltzmann Transport: Relaxation-Time Approximation}
%%%%%%%%

In the relaxation-time approximation 
of the Boltzmann transport\footnote{See, for example, 
Ziman: {\it Principles of the Theory of Solids}, 
2nd edition (Cambridge Univ. Press, Cambridge, 1972) 
for the Boltzmann transport of electrons. 
The Boltzmann transport of Cooper pairs is discussed 
in \S 3.7 of \cite{new}. } 
the expectation value of the charge current ${\bf J}^e$ 
and the heat current ${\bf J}^Q$ are given as\footnote{
Here ${\bf J}^e=(J^e_x, J^e_y, J^e_z)$, ${\bf J}^Q=(J^Q_x, J^Q_y, J^Q_z)$ 
and ${\bf v}_{\bf p}=(v_x, v_y, v_z)$. }
\begin{equation}
J_x^e = \big\langle e v_x \big\rangle 
\equiv 2 e \sum_{\bf p} v_x g_{\bf p}, 
\end{equation}
and 
\begin{equation}
J_x^Q = \big\langle \xi_{\bf p} v_x \big\rangle 
\equiv 2 \sum_{\bf p} \xi_{\bf p} v_x g_{\bf p} , 
\end{equation}
where 
$g_{\bf p}$ is the deviation of the distribution function\footnote{
Since we describe the distribution function 
of quasi-particles by the Boltzmann equation, 
$\xi_{\bf p}$ and $v_{\bf p}$ represent 
the energy and the velocity of an quasi-particle 
with a renormalized mass $m^*$ 
within the discussion of the Boltzmann transport. 
Namely, $\xi_{\bf p} = v_F \cdot (|{\bf p}| - p_F)$ 
and ${\bf v}_{\bf p}={\bf p}/m^*$ 
where $p_F$ is the Fermi momentum and $v_F = p_F / m^*$. 
To keep the notation simple, 
we do not distinguish $m^*$ and $m$ in the following. } 
from the equilibrium distribution function $f(\xi_{\bf p})$ 
defined by (\ref{FD}) and $e$ is the charge of an electron ($e<0$). 
Throughout the series of three notes 
the Zeeman splitting is neglected 
so that the spin degrees of freedom 
is accounted only by the degeneracy factor $2$. 

If a static electric field ${\bf E} = (E_x, 0, 0)$ is applied, 
the resulting deviation is 
\begin{equation}
g_{\bf p} = e E_x v_x \tau 
\Big( - { \partial f(\xi_{\bf p}) \over \partial \xi_{\bf p} } \Big), 
\end{equation}
so that 
\begin{equation}
J_x^e = 2 e^2 E_x \sum_{\bf p} v_x^2 \tau 
\Big( - { \partial f(\xi_{\bf p}) \over \partial \xi_{\bf p} } \Big) 
\equiv \sigma_{xx} E_x. 
\label{Drude-Bol} 
\end{equation}
In the same manner 
\begin{equation}
J_x^Q = 2 e E_x \sum_{\bf p} \xi_{\bf p} v_x^2 \tau 
\Big( - { \partial f(\xi_{\bf p}) \over \partial \xi_{\bf p} } \Big) 
\equiv \tilde\alpha_{xx} E_x. 
\label{Drude-Bol-Q} 
\end{equation}

If a static magnetic field ${\bf H}$ is applied additionally, 
the resulting ${\bf J}^e$ is 
\begin{equation}
{\bf J}^e = 2 e^2 \sum_{\bf p} 
{ v_x^2 \tau \over 1 + (\omega_c \tau)^2 }
\Big( - { \partial f(\xi_{\bf p}) \over \partial \xi_{\bf p} } \Big) 
\Big[ {\bf E} - { e \tau \over m } \big( {\bf H}\times{\bf E} \big) \Big], 
\end{equation}
where $\omega_c = |e{\bf H}|/m$. 

If a static temperature gradient ${\nabla T}$ is applied additionally, 
its effect is taken into account as 
\begin{equation}
e {\bf E}' = e {\bf E} - \xi_{\bf p} {\nabla T \over T}, 
\end{equation}
so that\footnote{
The treatment of the magnetic field in 
Reizer and Sergeev: Phys. Rev. B {\bf 61}, 7340 (2000) 
is not compatible with the Boltzmann transport. 
That in 
Sergeev, Reizer and Mitin: Phys. Rev. B {\bf 77}, 064501 (2008) 
is not compatible with the GL transport. 
In these works 
the gauge invariance is not satisfied in ordinary manner 
where the momentum is combined with the vector potential 
and the energy with the scalar potential. 
While the direction of the electron motion is affected, 
the value of the energy is not affected by the magnetic field. 
Some criticisms against these works have been made by 
Serbyn, Skvortsov and Varlamov: arXiv:1012.4316. 
However, 
Sergeev, Reizer and Mitin: arXiv:1101.4186 still claim their validity. } 
\begin{equation}
{\bf J}^e = 2 e^2 \sum_{\bf p} 
{ v_x^2 \tau \over 1 + (\omega_c \tau)^2 }
\Big( - { \partial f(\xi_{\bf p}) \over \partial \xi_{\bf p} } \Big) 
\Big[ {\bf E}' - { e \tau \over m } \big( {\bf H}\times{\bf E}' \big) \Big], 
\end{equation}
and 
\begin{equation}
{\bf J}^Q = 2 e \sum_{\bf p} 
{ v_x^2 \tau \over 1 + (\omega_c \tau)^2 } \, \xi_{\bf p} 
\Big( - { \partial f(\xi_{\bf p}) \over \partial \xi_{\bf p} } \Big) 
\Big[ {\bf E}' - { e \tau \over m } \big( {\bf H}\times{\bf E}' \big) \Big]. 
\label{J-Q-EB} 
\end{equation}

%%%%%%%%
\section{Boltzmann Transport: Exact Formula}
%%%%%%%%

The Boltzmann equation in linear order of ${\bf E}$ 
for an interacting electron system under static electromagnetic field 
is 
\begin{equation}
e{\bf E}\cdot{\bf v}_{\bf p} 
{ \partial f(\xi_{\bf p}) \over \partial \xi_{\bf p} } 
+ e \big( {\bf v}_{\bf p} \times {\bf H} \big)\cdot 
{ \partial g_{\bf p} \over \partial {\bf p} } 
= C_{\bf p}, \label{Bol} 
\end{equation}
where the collision term $C_{\bf p}$ is given as 
\begin{equation}
C_{\bf p} = 
\sum_{\bf p'} \Big\{ C_{\bf pp'}g_{\bf p'} - C_{\bf p'p}g_{\bf p} \Big\} 
\equiv - \sum_{\bf p'} \big( \tau_{\rm tr}^{-1} \big)_{\bf pp'}g_{\bf p'}, 
\end{equation}
with 
\begin{equation}
\big( \tau_{\rm tr}^{-1} \big)_{\bf pp'} = 
{ 1 \over \tau_{\bf p} }\delta_{\bf p,p'} - C_{\bf pp'}, 
\end{equation}
and 
\begin{equation}
{ 1 \over \tau_{\bf p} } \equiv \sum_{\bf p'} C_{\bf p'p}. 
\end{equation}
Here $C_{\bf p'p}$ is the transition rate from ${\bf p}$ to ${\bf p'}$ 
and has a symmetry $C_{\bf p'p} = C_{\bf pp'}$. 

This linear Boltzmann equation is formally solved exactly\footnote{
See, for example, Kotliar, Sengupta and Varma: 
Phys. Rev. B {\bf 53}, 3573 (1996). } 
as follows. 
By introducing the matrix $A$ 
\begin{equation}
A_{\bf pp'} = \big( \tau_{\rm tr}^{-1} \big)_{\bf pp'} 
- e \, \Big( {\bf v}_{\bf p} \times 
{ \partial \over \partial {\bf p} } \Big) 
\cdot {\bf H} \, \, \delta_{\bf p,p'}, 
\end{equation}
(\ref{Bol}) is written in the form 
\begin{equation}
\sum_{\bf p'} A_{\bf pp'} g_{\bf p'} = 
e{\bf E}\cdot{\bf v}_{\bf p} 
\Big( - { \partial f(\xi_{\bf p}) \over \partial \xi_{\bf p} } \Big). 
\end{equation}
If we get the inverse matrix $A^{-1}$, 
the deviation of the distribution is known as 
\begin{equation}
g_{\bf p'} = \sum_{\bf p} A_{\bf p'p}^{-1} 
e{\bf E}\cdot{\bf v}_{\bf p} 
\Big( - { \partial f(\xi_{\bf p}) \over \partial \xi_{\bf p} } \Big). 
\label{g(p)}
\end{equation}
Using the identity for operators P and Q 
\begin{equation}
\big( P-Q \big)^{-1} = 
P^{-1} + P^{-1} Q P^{-1} + P^{-1} Q P^{-1} Q P^{-1} + \cdot\cdot\cdot , 
\end{equation}
the inverse matrix is obtained 
as an expansion in terms of the magnetic field 
\begin{align}
A_{\bf p'p}^{-1} & = \big( \tau_{\rm tr} \big)_{\bf p'p} 
+ e \sum_{{\bf p}_1} \big( \tau_{\rm tr} \big)_{{\bf p}'{\bf p}_1} \! 
\bigg[ \! \Big( \! {\bf v}_{{\bf p}_1} \! \times 
\! { \partial \over \partial {\bf p}_1 } \! \Big) \! 
\cdot \! {\bf H} \bigg] \! 
\big( \tau_{\rm tr} \big)_{{\bf p}_1 {\bf p}} 
\nonumber \\ 
& + e^2 \! \sum_{{\bf p}_1} \! \sum_{{\bf p}_2} \! 
\big( \tau_{\rm tr} \big)_{{\bf p}'{\bf p}_1} \! 
\bigg[ \! \Big( \! {\bf v}_{{\bf p}_1} \! \times 
\! { \partial \over \partial {\bf p}_1 } \! \Big) \! 
\cdot \! {\bf H} \bigg] \! 
\big( \tau_{\rm tr} \big)_{{\bf p}_1 {\bf p}_2} \! 
\bigg[ \! \Big( \! {\bf v}_{{\bf p}_2} \! \times 
\! { \partial \over \partial {\bf p}_2 } \! \Big) \! 
\cdot \! {\bf H} \bigg] \! 
\big( \tau_{\rm tr}  \big)_{{\bf p}_2 {\bf p}} 
+ \cdot\cdot\cdot , \label{inverse-A}
\end{align}
Employing (\ref{g(p)}) and (\ref{inverse-A}) we can calculate 
\begin{equation}
{\bf J}^e = 2 e \sum_{\bf p'} {\bf v}_{\bf p'} g_{\bf p'}, 
\end{equation}
and the conductivity tensor $\sigma^{\mu\nu}$ is introduced\footnote{
Here ${\bf J}^e = (J^e_x, J^e_y, J^e_z)$, ${\bf E} = (E_x, E_y, E_z)$ 
and ${\bf v}_{\bf p} = (v_{\bf p}^x, v_{\bf p}^y, v_{\bf p}^z)$ 
with $\mu, \nu = x,y,z$. } as 
\begin{equation}
J^e_\mu = \sum_{\nu} \sigma^{\mu\nu} E_{\nu}. 
\end{equation}
Thus the conductivity tensor is expressed as 
\begin{equation}
\sigma^{\mu\nu} = 2 e^2 \sum_{\bf p'} \sum_{\bf p} 
v_{\bf p'}^\mu A_{\bf p'p}^{-1} v_{\bf p}^\nu 
\Big( - { \partial f(\xi_{\bf p}) \over \partial \xi_{\bf p} } \Big), 
\end{equation}
and this expression is microscopically derived 
on the basis of the Fermi-liquid theory.\footnote{
The microscopic derivation of the conductivity is done by 
\'Eliashberg: Sov. Phys. JETP {\bf 14}, 886 (1962). 
This work is extended to the case of the Hall conductivity 
by Kohno and Yamada: Prog. Theor. Phys. {\bf 80}, 623 (1988) 
and to the magneto-conductivity 
by Kontani: Phys. Rev. B {\bf 64}, 054413 (2001). } 
The conductivity in the absence of magnetic field is obtained as 
\begin{equation}
\sigma^{xx} = 2 e^2 \sum_{\bf p'} \sum_{\bf p} 
v_{\bf p'}^x \big( \tau_{\rm tr} \big)_{\bf p'p} v_{\bf p}^x 
\Big( - { \partial f(\xi_{\bf p}) \over \partial \xi_{\bf p} } \Big). 
\end{equation}

%%%%%%%%
\section{Charge and Heat Currents}
%%%%%%%%

The charge current of electrons, 
created by $\psi_\sigma^\dag({\bf x})$ 
and annihilated by $\psi_\sigma({\bf x})$, is 
\begin{equation}
{\bf j}^e({\bf x}) = { e \over 2mi } \sum_\sigma 
\Big[ \psi_\sigma^\dag({\bf x}) \Big( \nabla \psi_\sigma({\bf x}) \Big) 
- \Big( \nabla \psi_\sigma^\dag({\bf x}) \Big) \psi_\sigma({\bf x}) \Big], 
\label{current-e} 
\end{equation}
and satisfies the conservation law 
\begin{equation}
{\dot \rho}({\bf x}) + \nabla \cdot {\bf j}^e({\bf x}) = 0, 
\label{con-je}
\end{equation}
where 
\begin{equation}
\rho({\bf x}) = e  \sum_\sigma 
\psi_\sigma^\dag({\bf x}) \psi_\sigma({\bf x}) 
\equiv  j^e_0({\bf x}), 
\end{equation}
is the charge density. 
Using the four-divergence (\ref{con-je}) is written as 
\begin{equation}
\sum_{\mu=0}^3 {\partial \over \partial x_\mu} j^e_\mu(x) = 0, 
\end{equation}
where $x=(t,{\bf x})=(x_0, x_1,x_2,x_3)$ with $\mu = 0,1,2,3$. 

The energy current\footnote{
Via the energy-momentum tensor the energy current is given as 
\begin{equation}
u_\mu = \sum_\sigma \bigg( 
{\partial {\cal L} \over \partial(\partial \psi_\sigma/\partial x_\mu)} 
\dot\psi_\sigma + \dot\psi_\sigma^\dag 
{\partial {\cal L} \over \partial(\partial \psi_\sigma^\dag/\partial x_\mu)}
\bigg), \nonumber 
\end{equation}
where $\mu = 1,2,3$. 
The Lagrangian density ${\cal L}$ 
can be replaced by its kinetic part 
\begin{equation}
{\cal L_{\rm kin}}=-{1 \over 2m}\sum_\sigma 
\nabla \psi_\sigma^\dag \cdot \nabla \psi_\sigma, 
\nonumber 
\end{equation}
because its interaction part ${\cal L_{\rm int}}$ 
does not contain the derivative of $\psi_\sigma^\dag$ and $\psi_\sigma$. 
See, for example, Langer: Phys. Rev. {\bf 128}, 110 (1962) 
or \S 10.3 in \cite{old}. } 
of electrons is 
\begin{equation}
{\bf u}({\bf x}) = - { 1 \over 2m } \sum_\sigma 
\Big[ \Big( \nabla \psi_\sigma^\dag({\bf x}) \Big) \dot\psi_\sigma({\bf x}) 
+ \dot\psi_\sigma^\dag({\bf x}) \Big( \nabla \psi_\sigma({\bf x}) \Big) \Big], 
\label{u_x} 
\end{equation}
and satisfies the conservation law 
\begin{equation}
{\dot h}({\bf x}) + \nabla \cdot {\bf u}({\bf x}) = 0, 
\end{equation}
where $h({\bf x})$ is the Hamiltonian density. 

According to the energy conservation $dU = dQ + \mu dN$, 
the heat current is 
\begin{equation}
{\bf j}^Q({\bf x}) = {\bf u}({\bf x}) - {\mu \over e}{\bf j}^e({\bf x}), 
\end{equation}
and satisfies the conservation law 
\begin{equation}
\sum_{\mu=0}^3 {\partial \over \partial x_\mu} j^Q_\mu(x) = 0, 
\end{equation}
where 
\begin{equation}
j^Q_0(x) = h({\bf x}) - {\mu \over e}\rho({\bf x}). 
\label{j^Q_0(x)} 
\end{equation}

%%%%%%%%
\section{Kubo Formula: Linear Response to Electric Field}
%%%%%%%%

We observe the expectation value of the charge current ${\bf J}^e$ 
caused by the external electric field ${\bf E}$ as 
\begin{equation}
J^e_\mu({\bf k}, \omega) = 
\sum_\nu \sigma_{\mu\nu}({\bf k}, \omega) E_\nu({\bf k}=0, \omega). 
\end{equation}
The conductivity tensor $\sigma$ is given 
by the Kubo formula\footnote{
I follow Fukuyama, Ebisawa and Wada: Prog. Theor. Phys. {\bf 42}, 494 (1969) 
concerning the use of the Kubo formula. 
The sign of $\Phi_{\mu\nu}^e$ is opposite 
to $Q_{\mu\nu}$ in \cite{new} and [AGD] and 
to $K_{\mu\nu}$ in \cite{Sch} and [FW]. 

[AGD] $\equiv$ Abrikosov, Gorkov and Dzyaloshinski: 
{\it Methods of Quantum Field Theory in Statistical Physics} 
(Dover, New York, 1975). 

[FW] $\equiv$ Fetter and Walecka: {\it Quantum Theory of 
Many-Particle Systems} (McGraw-Hill, New York, 1971). } 
(linear response theory) 
\begin{equation}
\Phi_{\mu\nu}^e({\bf k}, i\omega_\lambda) = 
\int_0^\beta d \tau \, e^{i\omega_\lambda\tau} 
\big\langle T_\tau \big\{ \, 
j^e_\mu({\bf k}; \tau) \, j^e_\nu({\bf k} \!=\! 0) \big\} \big\rangle, 
\label{def-Phi-e} 
\end{equation}
and 
\begin{equation}
\sigma_{\mu\nu}({\bf k}, \omega) = { 1 \over i\omega } 
\big[ \Phi_{\mu\nu}^e({\bf k}, \omega + i\delta) 
- \Phi_{\mu\nu}^e({\bf k}, i\delta) \big]. 
\end{equation}
Here 
\begin{equation}
j^e_\mu({\bf k}; \tau) = e^{K\tau} \, j^e_\mu({\bf k}) \,  e^{-K\tau}, 
\end{equation}
and 
\begin{equation}
{\bf j}^e({\bf k}) = \int d^3 x \, e^{-i{\bf k}\cdot{\bf x}} \, 
{\bf j}^e({\bf x}). 
\end{equation}
The Fourier transform of (\ref{current-e}) results in 
\begin{align}
{\bf j}^e({\bf k}) 
& = {e \over m} \sum_{\bf p} \Big( {\bf p} + { {\bf k} \over 2 }\Big) 
\big( a_{\bf p}^\dag a_{\bf p+k} + b_{\bf p}^\dag b_{\bf p+k} \big) 
\nonumber \\
& = e \sum_{\bf p} { 1 \over 2 } 
\big( {\bf v}_{\bf p} + {\bf v}_{\bf p+k} \big) 
\big( a_{\bf p}^\dag a_{\bf p+k} + b_{\bf p}^\dag b_{\bf p+k} \big), 
\label{j^e(k)} 
\end{align}
and in the limit of ${\bf k}\rightarrow 0$ 
\begin{equation}
{\bf j}^e({\bf k} \!=\! 0) = 
e \sum_{\bf p} {\bf v}_{\bf p} 
\big( a_{\bf p}^\dag a_{\bf p} + b_{\bf p}^\dag b_{\bf p} \big) 
= e \sum_{\bf p} \sum_\sigma {\bf v}_{\bf p} 
c_{{\bf p}\sigma}^\dag c_{{\bf p}\sigma}, 
\end{equation}
where $c_{{\bf p}\sigma}^\dag$ represents the creation operator 
of an electron with momentum ${\bf p}$ and spin $\sigma$ and 
$c_{{\bf p}\sigma}$ represents the annihilation operator 
($\sigma = \uparrow, \downarrow$) introduced by 
\begin{equation}
\psi_\sigma^\dag({\bf x}) = \sum_{\bf p} 
e^{-i{\bf p}\cdot{\bf x}} c_{{\bf p}\sigma}^\dag, 
\ \ \ \ \ \ \ \ \ \ 
\psi_\sigma({\bf x}) = \sum_{\bf p} 
e^{i{\bf p}\cdot{\bf x}} c_{{\bf p}\sigma}. 
\end{equation}

In the presence of the temperature gradient 
both ${\bf E}$ and $\nabla T$ are the causes of 
the observed charge current ${\bf J}^e$ and heat current ${\bf J}^Q$ 
as \cite{USH} 
\begin{equation}
\left( \begin{array}{c} {\bf J}^e \\ {\bf J}^Q \\ \end{array} \right) =
\left( \begin{array}{cc} \sigma & \alpha \\ 
                   \tilde\alpha & \kappa \\ \end{array} \right) 
\left( \begin{array}{c} {\bf E} \\ -\nabla T \\ \end{array} \right), 
\end{equation}
where $\kappa$ is the thermal conductivity tensor and 
$\alpha$ and $\tilde\alpha$ are thermo-electric tensors. 
Here $\tilde\alpha$ is obtained by the Kubo formula as follows 
and $\alpha$ is obtained via the Onsager relation\footnote{
We have regarded ${\bf E}$ and $-\nabla T/T$ as external forces 
and assumed the Onsager relation between the responses to these forces. } as 
\begin{equation}
\alpha = {1 \over T} \tilde\alpha. 
\label{Onsager} 
\end{equation}
The expectation value of the heat current ${\bf J}^Q$ 
caused by the external electric field ${\bf E}$ is 
\begin{equation}
J^Q_\mu({\bf k}, \omega) = 
\sum_\nu \tilde\alpha_{\mu\nu}({\bf k}, \omega) E_\nu({\bf k}=0, \omega), 
\end{equation}
and the thermo-electric tensor $\tilde\alpha$ is given 
by the Kubo formula 
\begin{equation}
\Phi_{\mu\nu}^Q({\bf k}, i\omega_\lambda) = 
\int_0^\beta d \tau \, e^{i\omega_\lambda\tau} 
\big\langle T_\tau \big\{ \, 
j^Q_\mu({\bf k}; \tau) \, j^e_\nu({\bf k} \!=\! 0) \big\} \big\rangle, 
\label{def-Phi-Q} 
\end{equation}
and 
\begin{equation}
\tilde\alpha_{\mu\nu}({\bf k}, \omega) = { 1 \over i\omega } 
\big[ \Phi_{\mu\nu}^Q({\bf k}, \omega + i\delta) 
- \Phi_{\mu\nu}^Q({\bf k}, i\delta) \big]. 
\end{equation}
Here 
\begin{equation}
j^Q_\mu({\bf k}; \tau) = e^{K\tau} \, j^Q_\mu({\bf k}) \,  e^{-K\tau}, 
\end{equation}
and 
\begin{equation}
{\bf j}^Q({\bf k}) = \int d^3 x \, e^{-i{\bf k}\cdot{\bf x}} \, 
{\bf j}^Q({\bf x}). 
\end{equation}

%%%%%%%%
\section{Jonson-Mahan Formula: Relation between Response Functions}
%%%%%%%%

Let us find the relation between the energy current 
\begin{equation}
u_x(t) = - { 1 \over 2m } \sum_\sigma 
\bigg( \dot\psi_\sigma^\dag(t) {\partial \psi_\sigma(t) \over \partial x} 
+ {\partial \psi_\sigma^\dag(t) \over \partial x} \dot\psi_\sigma(t) \bigg), 
\end{equation}
and the charge current 
\begin{equation}
j^e_x(t) = { e \over 2mi } \sum_\sigma 
\bigg( \psi_\sigma^\dag(t) {\partial \psi_\sigma(t) \over \partial x} 
- {\partial \psi_\sigma^\dag(t) \over \partial x} \psi_\sigma(t) \bigg). 
\end{equation}
Introducing a two-time function 
\begin{equation}
J_\sigma(t,t') = 
\bigg( \psi_\sigma^\dag(t) {\partial \psi_\sigma(t') \over \partial x} 
- {\partial \psi_\sigma^\dag(t) \over \partial x} \psi_\sigma(t') \bigg), 
\label{J2} 
\end{equation}
the energy current is written as\footnote{
To determine (\ref{u_x}) 
we have chosen one of the equivalent forms of the density ${\cal L}_{\rm kin}$ 
resulting in the same value 
after integration $L = \int d^3 x {\cal L}_{\rm kin}$. 
Here the equivalent forms of the current density is introduced 
via the integration by parts 
\begin{equation}
\int d^3 x 
\bigg( {\partial \over \partial t} - {\partial \over \partial t'} \bigg) 
J_\sigma(t,t') 
= 2 \int d^3 x 
\bigg( \dot\psi_\sigma^\dag(t) {\partial \psi_\sigma(t') \over \partial x} 
+ {\partial \psi_\sigma^\dag(t) \over \partial x} \dot\psi_\sigma(t') \bigg), 
\nonumber 
\end{equation}
where we have neglected the contribution of the surface. } 
\begin{equation}
u_x = - \lim_{t' \rightarrow t} {1 \over 2} 
\bigg( {\partial \over \partial t} - {\partial \over \partial t'} \bigg) 
{1 \over 2m} \sum_\sigma J_\sigma(t,t'). 
\label{u_x-t} 
\end{equation}
Since 
\begin{equation}
-i{\partial \over \partial t} = {\partial \over \partial\tau}, 
\end{equation}
we obtain 
\begin{equation}
j^Q_x = \lim_{\tau' \rightarrow \tau} {1 \over 2} 
\bigg( {\partial \over \partial \tau} - {\partial \over \partial \tau'} \bigg) 
{1 \over 2mi} \sum_\sigma J_\sigma(\tau,\tau'). 
\end{equation}
Thus $\Phi_{xx}^e$ is expressed as 
\begin{equation}
\Phi_{xx}^e({\bf k} \!=\! 0, i\omega_\lambda) = e \lim_{\tau'\rightarrow\tau} 
\int_0^\beta d \tau \, e^{i\omega_\lambda\tau} F(\tau,\tau'), 
\end{equation}
where we have introduced a function 
\begin{equation}
F(\tau,\tau') \equiv \sum_{\bf p} \sum_\sigma v_x 
\big\langle 
c_{{\bf p}\sigma}^\dag(\tau) c_{{\bf p}\sigma}(\tau') 
j^e_x({\bf k} \!=\! 0) \big\rangle, 
\end{equation}
via the Fourier transform of (\ref{J2}), with $v_x = p_x / m$. 
Here we have assumed that $\beta > \tau > \tau' > 0$ 
and dropped $T_\tau$ from (\ref{def-Phi-e}). 
Introducing a function 
\begin{equation}
S(\tau,\tau') = {1 \over 2} 
\bigg( {\partial \over \partial \tau} - {\partial \over \partial \tau'} \bigg) 
F(\tau,\tau'), 
\end{equation}
$\Phi_{xx}^Q$ is expressed as 
\begin{equation}
\Phi_{xx}^Q({\bf k} \!=\! 0, i\omega_\lambda) = \lim_{\tau'\rightarrow\tau} 
\int_0^\beta d \tau \, e^{i\omega_\lambda\tau} S(\tau,\tau'). 
\end{equation}

Employing the Fourier transform\footnote{
Neglecting the vertex correction 
\begin{equation}
F(\tau,\tau') = e \sum_{\bf pp'} \sum_{\sigma\sigma'} v_x 
\big\langle c_{{\bf p}\sigma}^\dag(\tau) c_{{\bf p}\sigma}(\tau') 
c_{{\bf p'}\sigma'}^\dag(0) c_{{\bf p'}\sigma'}(0) \big\rangle 
v'_x, 
\nonumber 
\end{equation}
is factorized as 
\begin{equation}
F(\tau,\tau') = e \sum_{\bf p} \sum_{\sigma} v_x^2 
\big\langle c_{{\bf p}\sigma}(\tau') c_{{\bf p}\sigma}^\dag(0) \big\rangle 
\big\langle c_{{\bf p}\sigma}^\dag(\tau)  c_{{\bf p}\sigma}(0) \big\rangle. 
\nonumber 
\end{equation}
Since 
\begin{equation}
\big\langle c_{{\bf p}\sigma}(\tau') c_{{\bf p}\sigma}^\dag(0) \big\rangle 
= - G({\bf p},\tau'), 
\ \ \ \ \ \ \ \ \ \ 
\big\langle c_{{\bf p}\sigma}^\dag(\tau)  c_{{\bf p}\sigma}(0) \big\rangle 
= G({\bf p}, - \tau), 
\nonumber 
\end{equation}
for $\tau, \tau' > 0$, 
\begin{equation}
F(\tau,\tau') = - e \sum_{\bf p} \sum_{\sigma} v_x^2 
G({\bf p},\tau') G({\bf p}, - \tau). 
\nonumber 
\end{equation}
Via the Fourier transform of the propagator 
\begin{equation}
G({\bf p}, \tau') = {1 \over \beta} \sum_{n'} 
G({\bf p}, i\varepsilon_{n'}) e^{-i\varepsilon_{n'} \tau'}, 
\ \ \ \ \ \ \ \ \ \ 
G({\bf p}, - \tau) = {1 \over \beta} \sum_n 
G({\bf p}, i\varepsilon_n) e^{i\varepsilon_n \tau}, 
\nonumber 
\end{equation}
$F(\tau,\tau')$ in this approximation 
is shown to have the form of (\ref{Fourier-F}). 

If we consider the vertex correction, we obtain 
\begin{equation}
F(\tau,\tau') = - \sum_{\bf p} \sum_{\sigma} v_x 
G({\bf p},\tau') G({\bf p}, - \tau) \tilde j^e_x, 
\nonumber 
\end{equation}
where $\tilde j^e_x$ is the renormalized current vertex 
so that (\ref{Fourier-F}) holds in general. } 
\begin{equation}
F(\tau,\tau') = {1 \over \beta^2} \sum_n \sum_{n'} 
F(i\varepsilon_n,i\varepsilon_{n'}) 
e^{i(\varepsilon_n\tau -\varepsilon_{n'}\tau')}, 
\label{Fourier-F} 
\end{equation}
with $\varepsilon_n$ and $\varepsilon_{n'}$ 
being the fermionic thermal frequency, 
$\Phi_{xx}^e$ is expressed as 
\begin{equation}
\Phi_{xx}^e({\bf k} \!=\! 0, i\omega_\lambda) = {e \over \beta} \sum_n 
F(i\varepsilon_n,i\varepsilon_n+i\omega_\lambda). 
\label{JM-e} 
\end{equation}
Since 
\begin{equation}
S(\tau,\tau') = {1 \over 2\beta^2} \sum_n \sum_{n'} 
(i\varepsilon_n + i\varepsilon_{n'}) 
F(i\varepsilon_n,i\varepsilon_{n'}) 
e^{i(\varepsilon_n\tau -\varepsilon_{n'}\tau')}, 
\end{equation}
we obtain\footnote{
We have employed 
\begin{equation}
\int_0^\beta d \tau \, e^{i\omega_\lambda\tau} 
e^{i(\varepsilon_n-\varepsilon_{n'})\tau} 
= \beta \delta_{\varepsilon_n + \omega_\lambda, \, \varepsilon_{n'}}. 
\nonumber 
\end{equation}
} 
\begin{equation}
\Phi_{xx}^Q({\bf k} \!=\! 0, i\omega_\lambda) = {1 \over \beta} \sum_n 
\Big( i\varepsilon_n + {i\omega_\lambda \over 2} \Big) 
F(i\varepsilon_n,i\varepsilon_n+i\omega_\lambda). 
\label{JM-Q} 
\end{equation}
Thus $\Phi_{xx}^e$ and $\Phi_{xx}^Q$ are related\footnote{
This relation is obtained by 
Jonson and Mahan: Phys. Rev. B {\bf 42}, 9350 (1990) 
and also discussed 
by Mahan: Solid State Physics {\bf 51}, 81 (1998). } 
by (\ref{JM-e}) and (\ref{JM-Q}) for electrons. 

If we regard the Cooper pair as the carrier\footnote{
Since $\mu=0$ for Cooper pairs, the heat and energy currents are 
identified, ${\bf j}^Q({\bf x}) = {\bf u}({\bf x})$, here. } 
of charge and heat, 
we can introduce phenomenologically the current operators as 
\begin{equation}
{\bf j}^e({\bf x}) = { e^* \over 2m^*i } 
\Big[ \Psi^\dag({\bf x}) \Big( \nabla \Psi({\bf x}) \Big) 
- \Big( \nabla \Psi^\dag({\bf x}) \Big) \Psi({\bf x}) \Big], 
\label{Cooper-je} 
\end{equation}
and 
\begin{equation}
{\bf j}^Q({\bf x}) = - { 1 \over 2m^* } 
\Big[ \Big( \nabla \Psi^\dag({\bf x}) \Big) \dot\Psi({\bf x}) 
+ \dot\Psi^\dag({\bf x}) \Big( \nabla \Psi({\bf x}) \Big) \Big], 
\label{Cooper-jQ} 
\end{equation}
where 
\begin{equation}
\Psi^\dag({\bf x}) = \sum_{\bf q} 
e^{-i{\bf q}\cdot{\bf x}} P_{\bf q}^\dag, 
\ \ \ \ \ \ \ \ \ \ 
\Psi({\bf x}) = \sum_{\bf q} 
e^{i{\bf q}\cdot{\bf x}} P_{\bf q}. 
\end{equation}
These are the GL expressions of the currents \cite{USH}. 
The effective charge and mass of Cooper pairs are given as 
\begin{equation}
e^* = 2e, 
\ \ \ \ \ \ \ \ \ \ 
m^* = 2m. 
\end{equation}
Employing the Cooper-pair currents (\ref{Cooper-je}) and (\ref{Cooper-jQ}) 
instead of the electron currents (\ref{current-e}) and (\ref{u_x}), 
we obtain 
\begin{equation}
\Phi_{xx}^e({\bf k} \!=\! 0, i\omega_\lambda) = {e^* \over \beta} \sum_m 
\tilde F(i\omega_m,i\omega_m+i\omega_\lambda), 
\label{JM-e-Cooper} 
\end{equation}
and 
\begin{equation}
\Phi_{xx}^Q({\bf k} \!=\! 0, i\omega_\lambda) = {1 \over \beta} \sum_m 
\Big( i\omega_m + {i\omega_\lambda \over 2} \Big) 
\tilde F(i\omega_m,i\omega_m+i\omega_\lambda), 
\label{JM-Q-Cooper} 
\end{equation}
where 
\begin{equation}
\tilde F(\tau,\tau') \equiv \sum_{\bf q} v_x^* 
\big\langle 
P_{\bf q}^\dag(\tau) P_{\bf q}(\tau') 
j^e_x({\bf k} \!=\! 0) \big\rangle, 
\end{equation}
with 
$\omega_m$ being the bosonic thermal frequency and $v_x^* = q_x / m^*$. 
Thus $\Phi_{xx}^e$ and $\Phi_{xx}^Q$ are related 
by (\ref{JM-e-Cooper}) and (\ref{JM-Q-Cooper}) for Cooper pairs. 

%%%%%%%%
\section{Ward Identities: Exact Results}
%%%%%%%%

%%%%%%%%%%%%%%%%%%%%%%%%%%%%%%%%%%%%%%%%%%
%\vskip 12mm
\begin{figure}[htbp]
\begin{center}
%\hskip -0.6cm
\includegraphics[width=8.0cm]{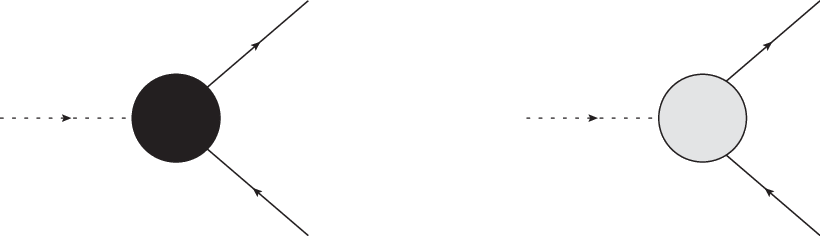}
\vskip 2mm
\caption{Current vertices for quasi-particles: 
The left black one is the charge current vertex. 
The right gray one is the heat current vertex. 
The broken line with arrow represents the coupling to the external field. }
\label{fig:Fig-Ward-QP}
\end{center}
\end{figure}
%%%%%%%%%%%%%%%%%%%%%%%%%%%%%%%%%%%%%%%%%%

\noindent
First 
we discuss the derivation~\cite{Sch} of the Ward identity 
for charge current vertex at zero temperature. 
We introduce the three-point function $\Lambda_\mu^e$, 
($\mu = 1,2,3,0$), defined by 
\begin{equation}
\Lambda_\mu^e(x,y,z) = \big\langle T 
\big\{ j_\mu^e(z) \psi_\uparrow(x) \psi_\uparrow^\dag(y) \big\} \big\rangle, 
\label{Lambda_mu^e(x,y,z)} 
\end{equation}
where 
$\langle A \rangle$ 
represents the expectation value of $A$ in the ground state, 
$T$ is the time-ordering operator and $z=({\bf z}, z_0)$ 
with coordinate vector ${\bf z}=(z_1,z_2,z_3)$ and time $z_0$. 
Here $\psi_\uparrow(x)$ and $\psi_\uparrow^\dag(y)$ are 
annihilation and creation operators of $\uparrow$-spin electron. 
The charge current $j_\mu^e$ of electrons 
obeys the charge-conservation law 
\begin{equation}
\sum_{\mu=0}^3 {\partial \over \partial z_\mu} j_\mu^e(z) = 0, \label{cc-law}
\end{equation}
where 
\begin{equation}
j_0^e(z) = 
e \psi_\uparrow^\dag(z) \psi_\uparrow(z) + 
e \psi_\downarrow^\dag(z) \psi_\downarrow(z). 
\end{equation}
The time ordering of three operators results in 
the summation of $3!$ terms as 
\begin{align}
\Lambda_\mu^e(x,y,z) &= 
\big\langle j_\mu^e(z) \psi_\uparrow(x) \psi_\uparrow^\dag(y) \big\rangle 
\theta(z_0-x_0)\theta(x_0-y_0) \nonumber \\ 
&- 
\big\langle j_\mu^e(z) \psi_\uparrow^\dag(y) \psi_\uparrow(x) \big\rangle 
\theta(z_0-y_0)\theta(y_0-x_0) \nonumber \\ 
&+ 
\big\langle \psi_\uparrow(x) j_\mu^e(z) \psi_\uparrow^\dag(y) \big\rangle 
\theta(x_0-z_0)\theta(z_0-y_0) \nonumber \\ 
&- 
\big\langle \psi_\uparrow^\dag(y) j_\mu^e(z) \psi_\uparrow(x) \big\rangle 
\theta(y_0-z_0)\theta(z_0-x_0) \nonumber \\ 
&+ 
\big\langle \psi_\uparrow(x) \psi_\uparrow^\dag(y) j_\mu^e(z) \big\rangle 
\theta(x_0-y_0)\theta(y_0-z_0) \nonumber \\ 
&- 
\big\langle \psi_\uparrow^\dag(y) \psi_\uparrow(x) j_\mu^e(z) \big\rangle 
\theta(y_0-x_0)\theta(x_0-z_0). 
\end{align}
Thus the time-derivative of $\Lambda_\mu^e$ results in 
\begin{align}
{\partial \over \partial z_0} \Lambda_0^e(x,y,z) = 
\delta(z_0-x_0) \Bigl( 
\theta(x_0-y_0) 
&\big\langle \big[ j_0^e(z), \psi_\uparrow(x) \big] 
\psi_\uparrow^\dag(y) \big\rangle 
\nonumber \\ 
- \theta(y_0-x_0) 
&\big\langle \psi_\uparrow^\dag(y) 
\big[ j_0^e(z), \psi_\uparrow(x) \big] \big\rangle 
\Bigr) \nonumber \\ 
+ 
\delta(z_0-y_0) \Bigl( 
\theta(x_0-y_0) 
&\big\langle \psi_\uparrow(x) 
\big[ j_0^e(z), \psi_\uparrow^\dag(y) \big] \big\rangle 
\nonumber \\ 
- \theta(y_0-x_0) 
&\big\langle \big[ j_0^e(z), \psi_\uparrow^\dag(y) \big] 
\psi_\uparrow(x) \big\rangle 
\Bigr) \nonumber \\ 
+ 
&\big\langle T 
\Big\{ {\partial j_0^e(z) \over \partial z_0} 
\psi_\uparrow(x) \psi_\uparrow^\dag(y) \Big\} \big\rangle . 
\label{del(t)-Lambda} 
\end{align}
Adding the divergence in terms of three coordinate variables 
and using again the time ordering operator $T$ 
we obtain the four-divergence of $\Lambda_\mu^e$ as 
\begin{align}
\sum_{\mu=0}^3 {\partial \over \partial z_\mu} \Lambda_\mu^e(x,y,z) = 
& \big\langle T 
\big\{ \big[ j_0^e(z), \psi_\uparrow(x) \big] 
\psi_\uparrow^\dag(y) \big\} \big\rangle 
\delta(z_0-x_0) \nonumber \\ 
+ 
& \big\langle T 
\big\{ \psi_\uparrow(x) 
\big[ j_0^e(z), \psi_\uparrow^\dag(y) \big] \big\} \big\rangle 
\delta(z_0-y_0) \nonumber \\ 
+ 
& \big\langle T 
\Big\{ \sum_{\mu=0}^3 {\partial j_\mu^e(z) \over \partial z_\mu} 
\psi_\uparrow(x) \psi_\uparrow^\dag(y) \Big\} \big\rangle. 
\label{divLambda} 
\end{align}
The last term on the right-hand side vanishes 
due to the charge-conservation law (\ref{cc-law}). 
Only equal space-time commutation relations are non-vanishing, 
\begin{equation}
\big[ j_0^e(z), \psi_\uparrow^\dag(y) \big] \delta(z_0-y_0) 
= e \psi_\uparrow^\dag(y) \delta^4(z-y), 
\end{equation}
and 
\begin{equation}
\big[ j_0^e(z), \psi_\uparrow(x) \big] \delta(z_0-x_0) 
= - e \psi_\uparrow(x) \delta^4(z-x), 
\end{equation}
so that the non-vanishing contribution becomes 
\begin{align}
\sum_{\mu=0}^3 {\partial \over \partial z_\mu} \Lambda_\mu^e(x,y,z) = 
& - e \big\langle T \big\{ \psi_\uparrow(x) 
\psi_\uparrow^\dag(y) \big\} \big\rangle \delta^4(z-x) 
\nonumber \\ 
& + e \big\langle T \big\{ \psi_\uparrow(x) 
\psi_\uparrow^\dag(y) \big\} \big\rangle \delta^4(z-y). 
\label{Ward-electron} 
\end{align}
Introducing the electron propagator $G(x,y)$ as 
\begin{equation}
G(x,y) = 
- i \big\langle T \big\{ \psi_\uparrow(x) 
\psi_\uparrow^\dag(y) \big\} \big\rangle, 
\label{G(t)}
\end{equation}
this relation is written into 
\begin{equation}
\sum_{\mu=0}^3 {\partial \over \partial z_\mu} \Lambda_\mu^e(x,y,z) = 
- i e G(x,y) \delta^4(z-x) + i e G(x,y) \delta^4(z-y). 
\label{LG} 
\end{equation}
Assuming the translational invariance 
we set $y=0$ and introduce the Fourier transform as 
\begin{equation}
\Lambda_\mu^e(p,k) = \int d^4 x e^{-ipx} \int d^4 z e^{-ikz} 
\big\langle T \big\{ j_\mu^e(z) 
\psi_\uparrow(x) \psi_\uparrow^\dag(0) \big\} \big\rangle, 
\end{equation}
where the four-momentum is defined as 
$p=({\bf p}, p_0)$ and $k=({\bf k}, k_0)$. 
The left-hand side of (\ref{LG}) is evaluated as 
\begin{equation}
\sum_{\mu=0}^3 {\partial \over \partial z_\mu} \Lambda_\mu^e(x,0,z) = 
\int {d^4 p \over (2\pi)^4} e^{ipx} \int {d^4 k \over (2\pi)^4}e^{ikz} 
\sum_{\mu=0}^3 i k_\mu \Lambda_\mu^e(p,k), 
\end{equation}
and the right-hand side is transformed as 
\begin{align}
\int d^4 x e^{-ipx} \int d^4 z e^{-ikz} 
\Bigl(
- G(x,0) \delta^4(z-x) + G(x,0) \delta^4(z) 
\Bigr) \nonumber \\ 
= - G(p+k) + G(p), 
\end{align}
where 
\begin{equation}
G(p) = 
\int d^4 x e^{-ipx} G(x,0). 
\end{equation}
Therefore we obtain 
\begin{equation}
\sum_{\mu=0}^3 k_\mu \Lambda_\mu^e(p,k)
= e G(p) - e G(p+k). 
\end{equation}
It should be noted that the factor $e$ represents 
the charge carried by an electron and 
is automatically taken into account by the commutation relation. 
The vertex function $\Gamma_\mu^e$ is introduced as 
\begin{equation}
\Lambda_\mu^e(p,k) = i G(p) \cdot \Gamma_\mu^e(p,k) \cdot i G(p+k), 
\label{LiG}
\end{equation}
in accordance with the definiton of the Green function (\ref{G(t)}). 
Then the Ward identity for the charge current vertex is obtained as 
\begin{equation}
\sum_{\mu=0}^3 k_\mu \Gamma_\mu^e(p,k) = 
e G(p)^{-1}- e G(p+k)^{-1}. 
\label{WI-e} 
\end{equation}
Since the Fourier transform is introduced as 
\begin{equation}
px = p_1x_1 + p_2x_2 + p_3x_3 - \epsilon t, 
\end{equation}
where $\epsilon$ is the energy and $t$ is the time, 
$x_0 = t$ and $p_0 = - \epsilon$ 
(and in the same manner $k_0 = - \omega$ 
with $\omega$ being the energy of the external field) 
in the zero-temperatute formalism. 

In the finite-temperature formalism 
we employ the time-ordering operator $T_\tau$ and 
consider the three-point function
\begin{equation}
\Lambda_\mu^e(x,y,z) = \big\langle T_\tau 
\big\{ j_\mu^e(z) \psi_\uparrow(x) \psi_\uparrow^\dag(y) \big\} \big\rangle, 
\end{equation}
where the real time $z_0$ and the imaginary time $\tau_z$ 
is related by $\tau_z = i z_0$ and 
$\langle A \rangle$ represents the expectation value of $A$ 
by grand canonical ensemble. 
Taking the charge-conservation law (\ref{cc-law}) 
into account we obtain\footnote{
The derivative $\partial\Lambda_0^e(x,y,z)/\partial\tau_z$ 
results in the same form as (\ref{del(t)-Lambda}) 
where, since $\partial / \partial \tau = - i \partial / \partial t$, 
the charge-conservation law becomes 
\begin{equation}
{\partial \rho({\bf x}) \over \partial \tau} 
- i \nabla \cdot {\bf j}^e({\bf x}) = 0. 
\nonumber 
\end{equation}
} 
\begin{align}
- i \sum_{\mu=0}^3 {\partial \over \partial z_\mu} \Lambda_\mu^e(x,y,z) = 
& \big\langle T_\tau 
\big\{ \big[ j_0^e(z), \psi_\uparrow(x) \big] 
\psi_\uparrow^\dag(y) \big\} \big\rangle \delta(\tau_z - \tau_x) 
\nonumber \\ 
+ 
& \big\langle T_\tau 
\big\{ \psi_\uparrow(x) 
\big[ j_0^e(z), \psi_\uparrow^\dag(y) \big] \big\} \big\rangle 
\delta(\tau_z - \tau_y). 
\label{finite-T-div} 
\end{align}
Here we have used the relation 
\begin{equation}
{\partial \over \partial \tau_z} = 
- i {\partial \over \partial z_0}. \label{del-tau}
\end{equation}
The finite-temperature vertex function $\Gamma_\mu^e$ is introduced as 
\begin{equation}
\Lambda_\mu^e(p,k) = 
\big[ -G(p) \big] \cdot \Gamma_\mu^e(p,k) \cdot \big[ -G(p+k) \big], 
\end{equation}
in accordance with the definiton of the thermal Green function 
\begin{equation}
G(x,y) = 
- \big\langle T_\tau \big\{ \psi_\uparrow(x) 
\psi_\uparrow^\dag(y) \big\} \big\rangle. 
\label{G(tau)} 
\end{equation}
Then the resulting Ward identity is the same form as (\ref{WI-e}) 
in the case of zero temperature. 
For the finite-temperature Ward identity 
the zeroth component of the four-momentum is 
$p_0 = - i \varepsilon_n$ with fermionic thermal frequency $\varepsilon_n$
and $k_0 = - i \omega_\lambda$ 
with bosonic thermal frequency $\omega_\lambda$. 

%%%%%%%%%%%%%%%%%%%%%%%%%%%%%%%%%%%%%%%%%%
\vskip 4mm
\begin{figure}[htbp]
\begin{center}
%\hskip -0.6cm
\includegraphics[width=8.0cm]{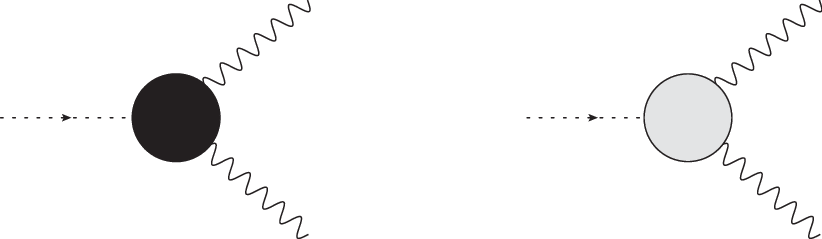}
\vskip 2mm
\caption{Current vertices for Cooper pairs. }
\label{fig:Fig-Ward-CP}
\end{center}
\end{figure}
%%%%%%%%%%%%%%%%%%%%%%%%%%%%%%%%%%%%%%%%%%

The extension\footnote{
I have derived the Ward identities for Cooper pairs with s-wave symmetry 
in arXiv:1108.0815. 
The Ward identities for anisotropic Cooper pairs have been discussed 
in arXiv:1108.5272. 
The Ward identities for Cooper pairs have been discussed 
in comparison with those for CDW and SDW in arXiv:1109.1404. 

[Nar] $\equiv$ arXiv:1108.0815, arXiv:1108.5272, arXiv:1109.1404. } 
of this Ward identity to the case of Cooper pairs 
is straightforward. 
In this note 
we only consider the local Cooper pair with s-wave symmetry 
in consistent with the local attractive interaction (\ref{local-g}). 
Replacing 
$\psi_\uparrow(x)$ by 
$\Psi(x) = \psi_\downarrow(x) \psi_\uparrow(x)$ and 
$\psi_\uparrow^\dag(y)$ by 
$\Psi^\dag(y) = \psi_\uparrow^\dag(y) \psi_\downarrow^\dag(y)$ 
in (\ref{Lambda_mu^e(x,y,z)}) 
we consider the three-point function $M_\mu^e$ as 
\begin{equation}
M_\mu^e(x,y,z) = \big\langle T 
\big\{ j_\mu^e(z) \Psi(x) \Psi^\dag(y) \big\} \big\rangle,
\end{equation}
where 
$\Psi(x)$ and $\Psi^\dag(y)$ are 
annihilation and creation operators of a Cooper pair 
which has a bosonic character. 
Using the commutation relation 
\begin{equation}
\big[ j_0^e(z), \Psi^\dag(y) \big] \delta(z_0-y_0)
= 2e \Psi^\dag(y) \delta^4(z-y), 
\end{equation}
and 
\begin{equation}
\big[ j_0^e(z), \Psi(x) \big] \delta(z_0-x_0)
= -2e \Psi(x) \delta^4(z-x), 
\end{equation}
the four-divergence of $M_\mu^e$ is expressed as 
\begin{align}
\sum_{\mu=0}^3 {\partial \over \partial z_\mu} M_\mu^e(x,y,z) = 
& - 2e \big\langle T \big\{ \Psi(x) 
\Psi^\dag(y) \big\} \big\rangle \delta^4(z-x) 
\nonumber \\ 
& + 2e \big\langle T \big\{ \Psi(x) 
\Psi^\dag(y) \big\} \big\rangle \delta^4(z-y), 
\end{align}
by repeating the same calculations as 
those for deriving (\ref{Ward-electron}). 
Here the difference between fermion and boson is handled 
solely by the time-ordering operator $T$ so that 
the expression of the divergence is common to fermion and boson. 
Introducing the Cooper-pair propagator $D(x,y)$ as 
\begin{equation}
D(x,y) = 
- i \big\langle T \big\{ \Psi(x) \Psi^\dag(y) \big\} \big\rangle, 
\end{equation}
we obtain the Ward identity 
\begin{equation}
\sum_{\mu=0}^3 k_\mu \Delta_\mu^e(q,k) = 
2e D(q)^{-1} - 2e D(q+k)^{-1}, 
\label{WIC-e} 
\end{equation}
for Cooper pairs 
where 
$\Delta_\mu^e$ is the counterpart of $\Gamma_\mu^e$ and 
$D(q)$ is the Fourier transform of $D(x,0)$ 
with four-momentum $q=({\bf q}, q_0)$. 
It should be noted that the factor $2e$ represents 
the charge carried by a Cooper pair and 
is automatically taken into account by the commutation relation. 

Although the above derivation for Cooper pairs 
is formulated at zero temperature, 
(\ref{WIC-e}) also holds at finite temperature 
with $q_0$ being a bosonic thermal frequency ($q_0 = - i \omega_m$). 
We are mainly interested in the normal metallic phase ($T>T_c$), 
the Cooper-pair propagator is a fluctuation propagator in this case. 

Next 
we discuss the derivation of the Ward identity 
for heat current vertex.\footnote{
See [Ono] $\equiv$ Ono: Prog. Theor. Phys. {\bf 46}, 757 (1971). } 
Here we only consider the local interaction.\footnote{
The derivation becomes very simple 
in the case of the local interaction [Kon]. 
See [Ono], [Kon], [Nar] 
for the derivation in the case of non-local interaction. 

[Kon] $\equiv$ Kontani: Phys. Rev. B {\bf 67}, 014408 (2003). }
We introduce the three-point function $\Lambda_\mu^Q$ 
defined by 
\begin{equation}
\Lambda_\mu^Q(x,y,z) = \big\langle T_\tau 
\big\{ j_\mu^Q(z) \psi_\uparrow(x) \psi_\uparrow^\dag(y) \big\} \big\rangle. 
\end{equation}
The heat current $j_\mu^Q$ obeys the conservation law 
\begin{equation}
\sum_{\mu=0}^3 {\partial \over \partial z_\mu} j_\mu^Q(z) = 0, 
\label{ec-law} 
\end{equation}
where $j_0^Q(z)$ is given by (\ref{j^Q_0(x)}) and is the density of (\ref{K}). 
The four-divergence of $\Lambda_\mu^Q$ at finite temperature becomes 
\begin{align}
- i \sum_{\mu=0}^3 {\partial \over \partial z_\mu} \Lambda_\mu^Q(x,y,z) = 
& \big\langle T_\tau 
\big\{ \big[ j_0^Q(z), \psi_\uparrow(x) \big] 
\psi_\uparrow^\dag(y) \big\} \big\rangle \delta(\tau_z - \tau_x) 
\nonumber \\ 
+ 
& \big\langle T_\tau 
\big\{ \psi_\uparrow(x) 
\big[ j_0^Q(z), \psi_\uparrow^\dag(y) \big] \big\} \big\rangle 
\delta(\tau_z - \tau_y), 
\label{divLamQ-h} 
\end{align}
as (\ref{finite-T-div}). 
Since the interaction among electrons is local, 
$ [j_0^Q(x), \psi_\uparrow(x)] = [K, \psi_\uparrow(x)] $ and 
$ [j_0^Q(y), \psi_\uparrow^\dag(y)] = [K, \psi_\uparrow^\dag(y)] $ 
so that using (\ref{Eq-of-motion}) 
\begin{equation}
\big[ j_0^Q(x), \psi_\uparrow(x) \big] = - i 
{\partial \over \partial x_0} \psi_\uparrow(x), \ \ \ 
\big[ j_0^Q(y), \psi_\uparrow^\dag(y) \big] = - i 
{\partial \over \partial y_0} \psi_\uparrow^\dag(y). 
\end{equation}
For simplicity of the representation of the Fourier transform, 
we discuss the zero-temperature case for a while. 
Employing (\ref{G(t)}) the four-divergence 
at zero temperature\footnote{
Starting from 
\begin{equation}
\Lambda_\mu^Q(x,y,z) = \big\langle T 
\big\{ j_\mu^Q(z) \psi_\uparrow(x) \psi_\uparrow^\dag(y) \big\} \big\rangle, 
\nonumber 
\end{equation}
we obtain 
\begin{align}
\sum_{\mu=0}^3 {\partial \over \partial z_\mu} \Lambda_\mu^Q(x,y,z) = 
& \big\langle T 
\big\{ \big[ j_0^Q(z), \psi_\uparrow(x) \big] 
\psi_\uparrow^\dag(y) \big\} \big\rangle 
\delta(z_0-x_0) \nonumber \\ 
+ 
& \big\langle T 
\big\{ \psi_\uparrow(x) 
\big[ j_0^Q(z), \psi_\uparrow^\dag(y) \big] \big\} \big\rangle 
\delta(z_0-y_0). 
\nonumber 
\end{align}
} is written as 
\begin{equation}
\sum_{\mu=0}^3 {\partial \over \partial z_\mu} \Lambda_\mu^Q(x,y,z) = 
{\partial \over \partial x_0} G(x,y) \delta^4(z-x) 
+ 
{\partial \over \partial y_0} G(x,y) \delta^4(z-y), 
\end{equation}
The Fourier transform of $\Lambda_\mu^Q$ becomes 
\begin{equation}
\int d^4 z e^{-ikz} \int d^4 x e^{-ip'x} \int d^4 y e^{ipy} 
\Lambda_\mu^Q(x,y,z) 
= \Lambda_\mu^Q(p,p-k) 
(2\pi)^4 \delta^4(-k-p'+p), 
\end{equation}
where we have assumed the translational invariance 
(regarded $\Lambda_\mu^Q$ as a function of $y-x$ and $z-x$) and set 
$ \Lambda_\mu^Q(x,y,z) = \Lambda_\mu^Q(x-y,z-x), $ and 
\begin{equation}
\Lambda_\mu^Q(p,p-k) = 
\int d^4 (x-y) e^{-ip(x-y)} \int d^4 (z-x) e^{-ik(z-x)} 
\Lambda_\mu^Q(x-y,z-x). 
\label{trans-inv-int} 
\end{equation}
At the same time we set $ G(x,y) = G(x\!-\!y) $ 
and introduce the Fourier transform 
\begin{equation}
{\partial \over \partial x_0} G(x-y) = 
\int {d^4 p' \over (2\pi)^4} e^{ip'(x-y)} i p'_0 G(p'). 
\end{equation}
Then we obtain 
\begin{equation}
\sum_{\mu=0}^3 k_\mu \Lambda_\mu^Q(p,p-k) = 
p_0 G(p) - (p_0 - k_0) G(p-k). 
\end{equation}
It should be noted that the factor $p_0$ or $p_0-k_0$ represents 
the energy carried by an electron and 
is automatically taken into account by the commutation relation. 
By shifting the four-momentum 
this relation is converted into the Ward identity 
\begin{equation}
\sum_{\mu=0}^3 k_\mu \Gamma_\mu^Q(p+k,p) = 
p_0 G(p+k)^{-1} - (p_0+k_0) G(p)^{-1}, 
\label{WI-Q} 
\end{equation}
for heat current vertex. 
Here we have used the same relation 
$ \Lambda_\mu^Q(p+k,p) = i G(p+k) \cdot \Gamma_\mu^Q(p+k,p) \cdot i G(p) $ 
as (\ref{LiG}). 
This identity also holds at finite temperature.\footnote{
As in the case of the Ward identity for charge-current vertex, 
the relation 
\begin{equation}
\Lambda_\mu^Q(p+k,p) = 
\big[ -G(p+k) \big] \cdot \Gamma_\mu^Q(p+k,p) \cdot \big[ -G(p) \big], 
\nonumber 
\end{equation}
should be taken into account. }

The extension\footnote{
See [Nar]. } of this Ward identity to the case of Cooper pairs 
is also straightforward. 
The Ward identity for heat current vertex of Cooper pairs is 
\begin{equation}
\sum_{\mu=0}^3 k_\mu \Delta_\mu^Q(q+k,q) = 
q_0 D(q+k)^{-1} - (q_0+k_0) D(q)^{-1}, 
\label{WIC-Q} 
\end{equation}
where 
$\Delta_\mu^Q$ is the counterpart of $\Gamma_\mu^Q$. 
It should be noted that the factor $q_0$ or $q_0+k_0$ represents 
the energy carried by a Cooper pair and 
is automatically taken into account by the commutation relation. 

Finally we discuss an application of the Ward identities. 
If we substitute the quasi-particle propagators\footnote{
We take the momenta of $G(p)^{-1}$ and $G(p+k)^{-1}$ symmetrically. } 
\begin{equation}
G(p)^{-1} = - p_0 - \xi_{{\bf p}-{{\bf k}\over 2}}, 
\ \ \ \ \ \ \ \ \ \ 
G(p+k)^{-1} = - p_0 - k_0 - \xi_{{\bf p}+{{\bf k}\over 2}}, 
\end{equation}
for the full propagators in (\ref{WI-e}), 
we obtain 
\begin{equation}
\sum_{\mu=0}^3 k_\mu \Gamma_\mu^e = e k_0 + 
e \big( \xi_{{\bf p}+{{\bf k}\over 2}} 
                  - \xi_{{\bf p}-{{\bf k}\over 2}} \big), 
\end{equation}
so that for $\mu=1,2,3$ 
\begin{equation}
\Gamma_\mu^e = {e \over m^*} p_\mu = e v_\mu, 
\label{Gamma^e-QP} 
\end{equation}
where we have employed (\ref{xi(p)}) 
by substituting the renormalized mass of quasi-particles $m^*$ for $m$. 
In the same manner 
we obtain 
\begin{equation}
\sum_{\mu=0}^3 k_\mu \Gamma_\mu^Q = 
- p_0 \xi_{{\bf p}+{{\bf k}\over 2}} 
+ (p_0+k_0) \xi_{{\bf p}-{{\bf k}\over 2}}, 
\end{equation}
from (\ref{WI-Q}). 
Thus for $\mu=1,2,3$ 
\begin{equation}
\Gamma_\mu^Q \fallingdotseq 
- {1 \over 2 m^*} p_\mu \big[ p_0 + (p_0+k_0) \big] 
= v_\mu \Big( i\varepsilon_n + {i\omega_\lambda \over 2} \Big). 
\label{Gamma^Q-QP} 
\end{equation}
These current vertices (\ref{Gamma^e-QP}) and (\ref{Gamma^Q-QP}) 
are consistent with (\ref{JM-e}) and (\ref{JM-Q}) 
of the Jonson-Mahan formula. 

If we substitute the Cooper-pair propagators\footnote{ In general 
this Ornstein-Zernike form is expected in the long-wavelength limit. 
We take the momenta of $D(q)^{-1}$ and $D(q+k)^{-1}$ symmetrically. } 
\begin{equation}
D({\bf q},i\omega_m\!=\!0)^{-1} = 
D(0,0)^{-1}\bigg[ 1 + \xi^2 \Big( {{\bf q}-{{\bf k}\over 2}} \Big)^2 \bigg], 
\label{static-D} 
\end{equation}
for the full propagators in (\ref{WIC-e}), 
we obtain 
\begin{equation}
\sum_{\mu=1}^3 k_\mu \Delta_\mu^e = 
2e D(0,0)^{-1} \xi^2 
\bigg[ \Big( {\bf q}-{{\bf k}\over 2} \Big)^2 
     - \Big( {\bf q}+{{\bf k}\over 2} \Big)^2 \bigg], 
\end{equation}
so that for $\mu=1,2,3$ 
\begin{equation}
\Delta_\mu^e = -4e D(0,0)^{-1} \xi^2 q_\mu. 
\label{Delta^e-CP} 
\end{equation}
In the same manner 
we obtain 
\begin{equation}
\Delta_\mu^Q = 
D(0,0)^{-1} \xi^2 q_\mu \big[ q_0 + (q_0+k_0) \big] 
= -2 D(0,0)^{-1} \xi^2 q_\mu 
\Big( i\omega_m + {i\omega_\lambda \over 2} \Big), 
\label{Delta^Q-CP} 
\end{equation}
from (\ref{WIC-Q}) for $\mu=1,2,3$. 
These current vertices (\ref{Delta^e-CP}) and (\ref{Delta^Q-CP}) 
are consistent with (\ref{JM-e-Cooper}) and (\ref{JM-Q-Cooper}) 
of the Jonson-Mahan formula. 

%%%%%%%%
\section{Quasi-particle Transport: Relaxation-Time Approximation}
%%%%%%%%

%%%%%%%%%%%%%%%%%%%%%%%%%%%%%%%%%%%%%%%%%%
%\vskip 12mm
\begin{figure}[htbp]
\begin{center}
%\hskip -0.6cm
\includegraphics[width=10.0cm]{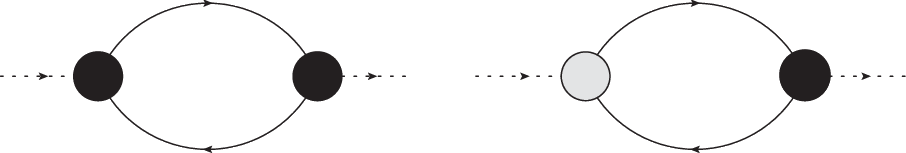}
\vskip 2mm
\caption{Quasi-particle transport 
within relaxation-time approximation: 
The left Feynman diagram describes 
the charge response to the external electric field. 
The right describes the heat response to the electric field. }
\label{fig:Fig-QP}
\end{center}
\end{figure}
%%%%%%%%%%%%%%%%%%%%%%%%%%%%%%%%%%%%%%%%%%

\noindent
Within the relaxation-time approximation 
we make a microscopic calculation 
equivalent to the Boltzmann transport in the following. 

The relaxation-time approximation corresponds to 
neglecting the vertex correction as shown in Fig.~4. 
In this approximation $\Phi_{xx}^e$ is given by\footnote{
The footnote for (\ref{Fourier-F}) shows 
how (\ref{relaxation-time approximation}) results from 
the Jonson-Mahan formula. 
Alternatively 
\begin{equation}
\Phi_{xx}^e({\bf k} \!=\! 0, i\omega_\lambda) = 
- {1 \over \beta} \sum_{\bf p} \sum_n \Gamma_x^e 
G(i\varepsilon_n) G(i\varepsilon_n+i\omega_\lambda)
\Gamma_x^e, 
\nonumber 
\end{equation}
results in (\ref{relaxation-time approximation}) 
using the Ward identity (\ref{Gamma^e-QP}). } 
\begin{equation}
\Phi_{xx}^e({\bf k} \!=\! 0, i\omega_\lambda) = 
- {2e^2 \over \beta} \sum_{\bf p} \sum_n v_x^2 
G(i\varepsilon_n) G(i\varepsilon_n+i\omega_\lambda), 
\label{relaxation-time approximation} 
\end{equation}
where the minus sign is the fermion-loop factor. 
We employ the analytic continuation 
of the thermal propagator $G(i\varepsilon_n)$ 
defined for pure imaginary frequencies 
to the retarded  or advanced propagator, $G^R(\epsilon)$ or $G^A(\epsilon)$, 
\begin{equation}
G^R(\epsilon) = {1 \over \epsilon - \xi_{\bf p} + i/2\tau}, 
\ \ \ \ \ \ \ \ \ \ 
G^A(\epsilon) = {1 \over \epsilon - \xi_{\bf p} - i/2\tau}, 
\end{equation}
for real frequencies. 
The analytic continuation results in $G^R(\epsilon)$ for $\varepsilon_n > 0$ 
and $G^A(\epsilon)$ for $\varepsilon_n < 0$. 
We have to calculate the discrete summation 
\begin{equation}
I^e(i\omega_\lambda) \equiv - {1 \over \beta} \sum_n 
G(i\varepsilon_n) G(i\varepsilon_n+i\omega_\lambda). 
\label{sum-GG} 
\end{equation}
%%%%%%%%%%%%%%%%%%%%%%%%%%%%%%%%%%%%%%%%%%
\begin{figure}[htbp]
\begin{center}
\hskip -0.3cm
\includegraphics[width=14cm]{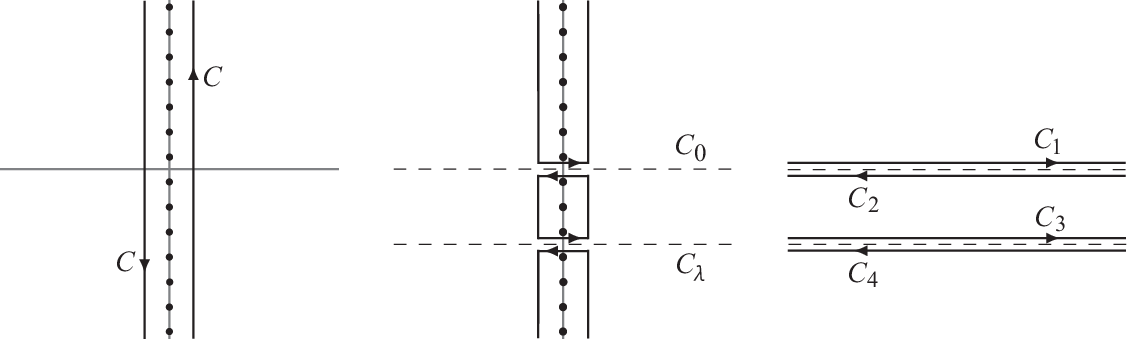}
\caption{Contours of the integral: 
These three panels are three different characterization 
of the same complex $z$-plane.
The solid line with an arrow represents the contour of the integral. 
The horizontal and vertical gray lines represent 
the real and imaginary axes of the complex $z$-plane. 
The dots on the imaginary axis represent the fermionic thermal frequencies. 
The broken lines $C_0$ and $C_\lambda$ represent the cuts along 
${\rm Im}\, z = 0$ and ${\rm Im}\, z = - \omega_\lambda$. }
\label{fig:contour}
\end{center}
\end{figure}
%%%%%%%%%%%%%%%%%%%%%%%%%%%%%%%%%%%%%%%%%%
The summation is transformed into the integral\footnote{
See, for example, \S 25 and Problem 7.6 in [FW]. 
Since $\tanh (z/2T) = 1 - 2 f(z)$, 
\begin{equation}
- I^e(i\omega_\lambda) = \int_C {dz \over 4\pi i} \tanh{z \over 2T} 
G(z)G(z+i\omega_\lambda). 
\nonumber 
\end{equation}
} on the contour $C$ in Fig.~\ref{fig:contour} as 
\begin{equation}
I^e(i\omega_\lambda)
= \int_C {dz \over 2\pi i} f(z) 
G(z)G(z+i\omega_\lambda). 
\label{sum2int} 
\end{equation}
The contour $C$ is transformed into $C_1$, $C_2$, $C_3$ and $C_4$ 
without encountering the singularity\footnote{
The propagator 
\begin{equation}
G(z) = \int_{-\infty}^\infty d \epsilon 
{\rho(\epsilon) \over z - \epsilon}, 
\nonumber 
\end{equation}
has a cut along ${\rm Im}\, z = 0$ so that 
$G(z)G(z+i\omega_\lambda)$ has cuts along 
${\rm Im}\, z = 0$ and ${\rm Im}\, z = - \omega_\lambda$. 
} of the propagator 
as shown in Fig.~\ref{fig:contour}. 
The analytic continuation of 
$ G(i\varepsilon_n) G(i\varepsilon_n+i\omega_\lambda) $ is 
$ G^R(z) G^R(z+i\omega_\lambda) $ on $C_1$, 
$ G^A(z) G^R(z+i\omega_\lambda) $ on $C_2$ and $C_3$ and 
$ G^A(z) G^A(z+i\omega_\lambda) $ on $C_4$. 
Taking into account of the direction of four contours we obtain 
\begin{align}
I^e(i\omega_\lambda)
& = \int_{C_0} {dz \over 2\pi i} f(z) 
G^R(z+i\omega_\lambda) \Big[ G^R(z) - G^A(z) \Big] 
\nonumber \\ 
& +
\int_{C_\lambda} {dz \over 2\pi i} f(z) 
G^A(z) \Big[ G^R(z+i\omega_\lambda) - G^A(z+i\omega_\lambda) \Big]. 
\label{Int-e-C}
\end{align}
Shifting the variable in the second integral in (\ref{Int-e-C}) we obtain 
\begin{equation}
I^e(i\omega_\lambda)
= \int_{-\infty}^\infty {d\epsilon \over 2\pi i} f(\epsilon) 
\Big[ G^R(\epsilon+i\omega_\lambda) + G^A(\epsilon-i\omega_\lambda) \Big] 
\Big[ G^R(\epsilon) - G^A(\epsilon) \Big], 
\end{equation}
where we have used the relation 
$f(\epsilon - i\omega_\lambda) = f(\epsilon)$ 
led by $e^{-\beta i\omega_\lambda} = 1$ 
for bosonic frequency $\omega_\lambda$. 
Since 
\begin{equation}
G^R(\epsilon) - G^A(\epsilon) = 2 i \, {\rm Im}G^R(\epsilon), 
\label{2iImG} 
\end{equation}
we finally obtain 
\begin{equation}
I^e(\omega+i\delta)
= \int_{-\infty}^\infty {d\epsilon \over \pi} f(\epsilon) 
\Big[ G^R(\epsilon + \omega) + G^A(\epsilon - \omega) \Big] 
{\rm Im}G^R(\epsilon). 
\label{Int-e-final} 
\end{equation}
To calculate the DC conductivity 
we only need the contribution linear in $\omega$ 
\begin{align}
I^e(\omega+i\delta) - I^e(i\delta) & \fallingdotseq 
\omega \int_{-\infty}^\infty {d\epsilon \over \pi} f(\epsilon) 
\bigg[ {\partial G^R(\epsilon) \over \partial \epsilon} 
     - {\partial G^A(\epsilon) \over \partial \epsilon} \bigg] 
{\rm Im}G^R(\epsilon) 
\nonumber \\ & = 
2i\omega \int_{-\infty}^\infty {d\epsilon \over \pi} f(\epsilon) 
\bigg[ {\partial \over \partial \epsilon} {\rm Im}G^R(\epsilon) \bigg] 
{\rm Im}G^R(\epsilon) 
\nonumber \\ & = 
i\omega \int_{-\infty}^\infty {d\epsilon \over \pi} 
\Big( - {\partial f(\epsilon) \over \partial \epsilon} \Big) 
\Big[ {\rm Im}G^R(\epsilon) \Big]^2, 
\label{I-e-omega} 
\end{align}
where we have employed the integration by parts. 
Thus the DC conductivity $\sigma_{xx}$ is given as 
\begin{align}
\sigma_{xx} & = \lim_{\omega \rightarrow 0} { 1 \over i\omega } 
\big[ \Phi_{xx}^e({\bf k} \!=\! 0, \omega + i\delta) 
- \Phi_{xx}^e({\bf k} \!=\! 0, i\delta) \big] 
\nonumber \\ 
& = 2 e^2 \int_{-\infty}^\infty {d\epsilon \over \pi} 
\Big( - {\partial f(\epsilon) \over \partial \epsilon} \Big) 
\sum_{\bf p} v_x^2 \Big[ {\rm Im}G^R(\epsilon) \Big]^2. 
\end{align}
If the life time $\tau$ is not so short 
($\epsilon_F\tau \gg 1$ where $\epsilon_F=p_F^2/2m$), 
${\rm Im}G^R(\epsilon)$ behaves as $-\pi\delta(\epsilon - \xi_{\bf p})$ 
so that\footnote{
See, for example, \S 8.1.2 in Mahan: 
{\it Many-Particle Physics}, 3rd edition 
(Kluwer Academic/Plenum, New York, 2000). } 
\begin{equation}
\sigma_{xx} \sim 2 e^2 \sum_{\bf p} 
\Big( - {\partial f(\xi_{\bf p}) \over \partial \xi_{\bf p}} \Big) v_x^2 
\int_{-\infty}^\infty {d\epsilon \over \pi} 
\Big[ {\rm Im}G^R(\epsilon) \Big]^2. 
\label{Drude-micro} 
\end{equation}
After the integration\footnote{
Using the definite integral 
\begin{equation}
\int_{-\infty}^\infty {dx \over (x^2+a^2)^2} = {\pi \over 2a^3}, 
\nonumber 
\end{equation}
for $a>0$ we obtain 
\begin{equation}
\int_{-\infty}^\infty d\epsilon 
\Big[ {\rm Im}G^R(\epsilon) \Big]^2 
= \int_{-\infty}^\infty d\epsilon 
\bigg( { 1/2\tau \over (\epsilon - \xi_{\bf p})^2 +(1/2\tau)^2 } \bigg)^2 
= \pi \tau. 
\nonumber 
\end{equation}
} over $\epsilon$ we obtain 
\begin{equation}
\sigma_{xx} \sim 2 e^2 \sum_{\bf p} 
\Big( - {\partial f(\xi_{\bf p}) \over \partial \xi_{\bf p}} \Big) 
v_x^2 \tau, 
\label{sigma-RTA} 
\end{equation}
which is equivalent to (\ref{Drude-Bol}). 

In the same manner\footnote{
Using (\ref{Gamma^e-QP}) and (\ref{Gamma^Q-QP}) 
\begin{equation}
\Phi_{xx}^Q({\bf k} \!=\! 0, i\omega_\lambda) = 
- {1 \over \beta} \sum_{\bf p} \sum_n \Gamma_x^Q 
G(i\varepsilon_n) G(i\varepsilon_n+i\omega_\lambda)
\Gamma_x^e, 
\nonumber 
\end{equation}
results in (\ref{relaxation-time approximation-Q}). } 
as (\ref{relaxation-time approximation}) 
$\Phi_{xx}^Q$ is given by 
\begin{equation}
\Phi_{xx}^Q({\bf k} \!=\! 0, i\omega_\lambda) = 
- {2e \over \beta} \sum_{\bf p} \sum_n v_x^2 
\Big( i\varepsilon_n + {i\omega_\lambda \over 2} \Big) 
G(i\varepsilon_n) G(i\varepsilon_n+i\omega_\lambda). 
\label{relaxation-time approximation-Q} 
\end{equation}
The discrete summation 
\begin{equation}
I^Q(i\omega_\lambda) \equiv - {1 \over \beta} \sum_n 
\Big( i\varepsilon_n + {i\omega_\lambda \over 2} \Big) 
G(i\varepsilon_n) G(i\varepsilon_n+i\omega_\lambda), 
\end{equation}
is transformed into the integral 
\begin{equation}
I^Q(\omega+i\delta) 
= \int_{-\infty}^\infty {d\epsilon \over \pi} f(\epsilon) 
\Big[ \Big( \epsilon + {\omega \over 2} \Big) 
G^R(\epsilon + \omega) 
+ \Big( \epsilon - {\omega \over 2} \Big) 
G^A(\epsilon - \omega) \Big] 
{\rm Im}G^R(\epsilon), 
\end{equation}
in the same manner as (\ref{Int-e-final}). 
The $\omega$-linear contribution is evaluated\footnote{
Using (\ref{2iImG}), the right-hand side 
of the first line of (\ref{Int-Q-linear}) is equal to 
\begin{equation}
i\omega \int_{-\infty}^\infty {d\epsilon \over \pi} f(\epsilon) 
\bigg[ 2 \epsilon {\partial \over \partial \epsilon} 
{\rm Im} G^R(\epsilon) + {\rm Im}G^R(\epsilon) \bigg] 
{\rm Im}G^R(\epsilon) 
= i\omega \int_{-\infty}^\infty {d\epsilon \over \pi} 
\Big\{ - 
\Big[ {\partial \over \partial \epsilon} 
\big( f(\epsilon)\cdot\epsilon \big) \Big] 
+ f(\epsilon) \Big\} \Big[ {\rm Im}G^R(\epsilon) \Big]^2, 
\nonumber 
\end{equation}
where we have employed the integration by parts. } as 
\begin{align}
I^Q(\omega+i\delta) - I^Q(i\delta) & \fallingdotseq 
\omega \int_{-\infty}^\infty {d\epsilon \over \pi} f(\epsilon) 
\bigg\{ \epsilon \bigg[  
  {\partial G^R(\epsilon) \over \partial \epsilon} 
- {\partial G^A(\epsilon) \over \partial \epsilon} \bigg] 
+ {1 \over 2} \Big[ G^R(\epsilon) - G^A(\epsilon) \Big] \bigg\} 
{\rm Im}G^R(\epsilon) 
\nonumber \\ & = 
i\omega \int_{-\infty}^\infty {d\epsilon \over \pi} 
\Big( - {\partial f(\epsilon) \over \partial \epsilon} \Big) 
\epsilon \Big[ {\rm Im}G^R(\epsilon) \Big]^2. 
\label{Int-Q-linear} 
\end{align}
Thus 
\begin{align}
\tilde\alpha_{xx}({\bf k} \!=\! 0, \omega \!\rightarrow\! 0) 
& = \lim_{\omega \rightarrow 0} { 1 \over i\omega } 
\big[ \Phi_{xx}^Q({\bf k} \!=\! 0, \omega + i\delta) 
- \Phi_{xx}^Q({\bf k} \!=\! 0, i\delta) \big] 
\nonumber \\ 
& = 2 e \int_{-\infty}^\infty {d\epsilon \over \pi} 
\Big( - {\partial f(\epsilon) \over \partial \epsilon} \Big) \epsilon 
\sum_{\bf p} v_x^2 \Big[ {\rm Im}G^R(\epsilon) \Big]^2. 
\end{align}
In the same manner as (\ref{Drude-micro}) we obtain\footnote{
Formally $\tilde\alpha_{xx}$ is proportional to the charge of electrons. 
However, the sign of the summation at low temperature 
is determined by that of $N_1$ 
where the density of states $N(\xi_{\bf p})$ is expanded as 
$ N(\xi_{\bf p}) \fallingdotseq N(0) + N_1 \cdot \xi_{\bf p}$. 
In the presence of particle-hole symmetry, 
$N_1$ and thus $\tilde\alpha_{xx}$ vanish. 
See (10.19) in \cite{old} for the discussion on the symmetry. } 
\begin{equation}
\tilde\alpha_{xx}({\bf k} \!=\! 0, \omega \!\rightarrow\! 0) 
\sim 2 e \sum_{\bf p} 
\Big( - {\partial f(\xi_{\bf p}) \over \partial \xi_{\bf p}} \Big) 
v_x^2 \xi_{\bf p} \tau, 
\end{equation}
which is equivalent to (\ref{Drude-Bol-Q}). 

As another way of evaluating the summation in (\ref{sum-GG}) 
we employ the spectral representation (\ref{sr-G}) and obtain\footnote{
Here the relation 
\begin{equation}
\lim_{\delta \rightarrow 0} {1 \over \beta} \sum_n 
{e^{i\varepsilon_n\delta} \over i\varepsilon_n - x} = f(x), \nonumber 
\end{equation}
is employed taking into account 
that $\varepsilon_n + \omega_\lambda$ is also a fermionic thermal frequency. 
See, for example, \S 25 in [FW]. } 
\begin{align}
I^e(i\omega_\lambda) & = - {1 \over \beta} \sum_n 
\int_{-\infty}^\infty 
{\rho(x) dx \over i\varepsilon_n - x} 
\int_{-\infty}^\infty 
{\rho(y) dy \over i\varepsilon_n + i\omega_\lambda - y} \nonumber \\ 
& = - \int_{-\infty}^\infty dx \int_{-\infty}^\infty dy 
{\rho(x) \rho(y) \over x - y + i\omega_\lambda} 
{1 \over \beta} \sum_n 
\bigg[ {1 \over i\varepsilon_n - x} 
- {1 \over i\varepsilon_n + i\omega_\lambda - y} \bigg] \nonumber \\ 
& = \int_{-\infty}^\infty dx \int_{-\infty}^\infty dy 
{\rho(x) \rho(y) \over x - y + i\omega_\lambda} 
\big[ f(y) - f(x) \big]. 
\end{align}
The imaginary part is 
\begin{align}
{\rm Im} \big[ I^e(\omega+i\delta) \big] 
& = - \pi \int_{-\infty}^\infty dx \int_{-\infty}^\infty dy \rho(x) \rho(y) 
\big[ f(y) - f(x) \big] \delta(x-y+\omega) \nonumber \\ 
& = - \pi \int_{-\infty}^\infty dx \rho(x) \rho(x+\omega) 
\big[ f(x+\omega) - f(x) \big] \nonumber \\ 
& = - \pi \int_{-\infty}^\infty dx \rho(x) 
\big[ \rho(x-\omega) - \rho(x+\omega)  \big] f(x). 
\end{align}
The $\omega$-linear contribution is estimated as 
\begin{equation}
{\rm Im} \big[ I^e(\omega+i\delta) \big] \fallingdotseq 
\pi \int_{-\infty}^\infty dx \rho(x) 
{\partial \rho(x) \over \partial x}\cdot 2\omega \cdot f(x). 
\end{equation}
Employing the integration by parts 
we obtain the final result equivalent to (\ref{I-e-omega}) 
\begin{equation}
{\rm Im} \big[ I^e(\omega+i\delta) \big] \fallingdotseq 
\pi \omega \int_{-\infty}^\infty dx \Big[ \rho(x) \Big]^2 
\Big( - {\partial f(x) \over \partial x} \Big). 
\end{equation}

%%%%%%%%
\section{GL Transport: Gaussian Fluctuation}
%%%%%%%%

Here we discuss the GL transport.\footnote{
We follow the discussion for the conductivity by Skocpol and Tinkham: 
Rep. Prog. Phys. {\bf 38}, 1049 (1975) and 
Tinkham: {\it Introduction to Superconductivity}, 2nd edition 
(McGraw-Hill, New York, 1996). Then we extend the discussion to the case of 
the thermo-electric tensor and our result (\ref{alpha-xx--GL}) is identical 
to that in \cite{new} and \cite{USH}. } 
We consider the GL free-energy density 
\begin{equation}
E({\bf x}) = \alpha \big|\Psi({\bf x})\big|^2
+ {1 \over 2m^*}\bigg| {\nabla \over i} \Psi({\bf x})\bigg|^2, 
\end{equation}
in the Gaussian approximation. 
Introducing the Fourier transform of the complex order-parameter field 
\begin{equation}
\Psi({\bf x}) = \sum_{\bf q} 
e^{i{\bf q}\cdot{\bf x}} \Psi_{\bf q}, 
\ \ \ \ \ \ \ \ \ \ 
\Psi^*({\bf x}) = \sum_{\bf q} 
e^{-i{\bf q}\cdot{\bf x}} \Psi_{\bf q}^*, 
\label{OP} 
\end{equation}
the GL free-energy $E$ is given by 
\begin{equation}
E = \int d^3x E({\bf x})
= \sum_{\bf q} \Big( \alpha + {q^2 \over 2m^*} \Big) 
\Psi_{\bf q}^* \Psi_{\bf q}. 
\end{equation}
The Gaussian fluctuation is evaluated as\footnote{
The Gauss integral for complex variable $z$ is performed as 
\begin{equation}
\int{dz^* dz \over 2 \pi i} e^{- a z^* z} = 
\int{du dv \over \pi} e^{- a (u^2 + v^2)} = {1 \over a}. 
\nonumber 
\end{equation}
} 
\begin{equation}
\big\langle \Psi_{\bf q}^* \Psi_{\bf q} \big\rangle 
={T \over \alpha + q^2 / 2m^*} 
={T \over \alpha} {1 \over 1 + \xi^2 q^2}, 
\label{ev-Gauss} 
\end{equation}
where $\langle A \rangle$ is the expectation value of $A$ 
under the Boltzmann weight $\exp(-E/T)$. 
Here the correlation length $\xi$ of the fluctuation is introduced as 
\begin{equation}
\xi^2 = {1 \over 2 m^* \alpha}. 
\label{corr-length} 
\end{equation}
The relaxation of the fluctuation is determined 
by the time-dependent GL equation 
\begin{equation}
\gamma {\partial \Psi_{\bf q} \over \partial t} 
= - {\delta E \over \delta \Psi_{\bf q}^*}. 
\label{TDGL} 
\end{equation}
In terms of the relaxation time $\tau_{\bf q}$ 
\begin{equation}
{1 \over \tau_{\bf q}} 
= {1 \over \gamma} \Big( \alpha + {q^2 \over 2m^*} \Big) 
= {\alpha \over \gamma} \big( 1 + \xi^2 q^2 \big), 
\end{equation}
(\ref{TDGL}) is written as 
\begin{equation}
{\partial \Psi_{\bf q} \over \partial t} 
= - {1 \over \tau_{\bf q}} \Psi_{\bf q}. 
\label{TDGL-tau} 
\end{equation}

The Kubo formula for classical variables is 
\begin{equation}
\sigma_{xx} = {1 \over T} \int_0^\infty 
\big\langle J_x^e(0) J_x^e(t) \big\rangle dt, 
\label{sigma-Kubo} 
\end{equation}
where the charge current\footnote{
The charge and heat currents corresponding to 
(\ref{Cooper-je}) and (\ref{Cooper-jQ}) are given by 
\begin{equation}
{\bf j}^e({\bf x}) = { e^* \over 2m^*i } 
\Big[ \Psi^*({\bf x}) \Big( \nabla \Psi({\bf x}) \Big) 
- \Big( \nabla \Psi^*({\bf x}) \Big) \Psi({\bf x}) \Big], 
\nonumber 
\end{equation}
and 
\begin{equation}
{\bf j}^Q({\bf x}) = - { 1 \over 2m^* } 
\Big[\Big( \nabla \Psi^*({\bf x}) \Big)\dot\Psi({\bf x}) + 
\dot\Psi^*({\bf x}) \Big( \nabla \Psi({\bf x}) \Big)\Big], 
\nonumber 
\end{equation}
as (4) and (5) in \cite{USH}. 
Using (\ref{OP}) the Fourier-transformed charge current is  
\begin{equation}
{\bf j}^e({\bf k}) = {e^* \over m^*} \sum_{\bf q} 
\Big( {\bf q} + {{\bf k} \over 2} \Big) \Psi_{\bf q}^* \Psi_{\bf q+k}, 
\nonumber 
\end{equation}
as (\ref{j^e(k)}). 
Thus the $x$-component of the uniform current $J_x^e=j_x^e({\bf k}\!=\!0)$ 
is given by (\ref{J_x(t)}). } 
is given by 
\begin{equation}
J_x^e(t) = {e^* \over m^*} \sum_{\bf q} q_x 
\Psi_{\bf q}^*(t) \Psi_{\bf q}(t). 
\label{J_x(t)} 
\end{equation}
The current-current correlation 
\begin{equation}
\big\langle J_x^e(0) J_x^e(t) \big\rangle 
= \Big({e^* \over m^*}\Big)^2 \sum_{\bf q} q_x^2 
\big\langle |\Psi_{\bf q}^*(0) \Psi_{\bf q}(t)|^2 \big\rangle, 
\end{equation}
is evaluated by using 
\begin{equation}
\big\langle |\Psi_{\bf q}^*(0) \Psi_{\bf q}(t)|^2 \big\rangle 
= \big\langle \Psi_{\bf q}^* \Psi_{\bf q} \big\rangle^2 
e^{-2t/\tau_{\bf q}}, 
\end{equation}
where the time-dependence of the order-parameter field 
is determined by (\ref{TDGL-tau}) as 
\begin{equation}
\Psi_{\bf q}(t) = \Psi_{\bf q} e^{-t/\tau_{\bf q}}. 
\label{TD-Psi} 
\end{equation}
After the time-integral we obtain 
\begin{equation}
\sigma_{xx} = 
\Big({e^* \over m^*}\Big)^2 {1 \over T} \sum_{\bf q} q_x^2 
\big\langle \Psi_{\bf q}^* \Psi_{\bf q} \big\rangle^2 
{\tau_{\bf q} \over 2}. 
\label{Drude-GL} 
\end{equation}
Using (\ref{ev-Gauss}) we obtain the DC conductivity 
\begin{equation}
\sigma_{xx} = \pi e^2 {T \over T-T_c} 
\sum_{\bf q} { \xi^4 q_x^2 \over (1 + \xi^2 q^2)^3 }. 
\end{equation}

In 2D the ${\bf q}$-summation is performed as\footnote{
Here we consider the film whose thickness is $d$ 
and assume $d \ll \xi$ so that 
\begin{equation}
\sum_{\bf q} \rightarrow 
{1 \over 2 \pi d} \int_0^\infty q dq. 
\nonumber 
\end{equation}
The replacement $q_x^2 \rightarrow q^2/2$ can be done 
for the present symmetric case in 2D. 
Since 
\begin{equation}
\int dx {x \over (x+1)^3} 
= \int dx {1 \over (x+1)^2} - \int dx {1 \over (x+1)^3}, 
\nonumber 
\end{equation}
we obtain 
\begin{equation}
\int_{0}^\infty dx {x \over (x+1)^3} 
= {1 \over 2}. 
\nonumber 
\end{equation}
} 
\begin{equation}
\sum_{\bf q} { \xi^4 q_x^2 \over (1 + \xi^2 q^2)^3 } 
= {1 \over 8 \pi d} \int_0^\infty dx {x \over (x + 1)^3} 
= {1 \over 16 \pi d}, 
\label{int-AL} 
\end{equation}
so that 
\begin{equation}
\sigma_{xx} = {e^2 \over 16 d} {T \over T-T_c}. 
\label{sigma-2D} 
\end{equation}

Next we consider the Kubo formula 
for the thermo-electric tensor $\tilde\alpha$ 
\begin{equation}
\tilde\alpha_{xx} = {1 \over T} \int_0^\infty 
\big\langle J_x^e(0) J_x^Q(t) \big\rangle dt, 
\label{alpha-Kubo} 
\end{equation}
where the heat current is given by 
\begin{equation}
J_x^Q(t) = - {i \over 2m^*} \lim_{t' \rightarrow t} 
\Big( {\partial \over \partial t} - {\partial \over \partial t'} \Big) 
\sum_{\bf q} q_x \Psi_{\bf q}^*(t) \Psi_{\bf q}(t'), 
\end{equation}
which corresponds to (\ref{u_x-t}). 
Although in (\ref{TD-Psi}) $\tau_{\bf q}$ is assumed to be real,\footnote{
If $\tau_{\bf q}$ is real, $\alpha_{xx}$ vanishes. 
Since holes carry opposite charge to electrons, 
$\alpha_{xx}$ and $\sigma_{xy}$ 
that are odd in the electric charge vanish by the cancelation 
between contributions from particle- and hole- charge currents 
in the presence of the particle-hole symmetry. 
On the other hand, such a cancelation does not work for the heat current. 
Even in the presence of the particle-hole symmetry 
$\sigma_{xx}$ and $\alpha_{xy}$ that are even in the electric charge 
do not vanish. 
See, for example, 
Niven and Smith: Phys. Rev. B {\bf 66}, 214505 (2002) 
for the consideration in terms of the symmetry. } 
here it is extended to be complex\footnote{
In general the complex GL equation 
\begin{equation}
{\partial \tilde\Psi \over \partial t} = (1+ic_0)\tilde\Psi 
+ (1+ic_1)\nabla^2 \tilde\Psi - (1+ic_2) \big| \tilde\Psi \big|^2 \tilde\Psi, 
\nonumber 
\end{equation}
is employed to study the dynamics of the complex order parameter $\tilde\Psi$. 
The origins of the imaginary part of $\tau_{\bf q}$ are discussed 
in \S 9.3 of \cite{new} for example. 
The non-vanishing imaginary part is related 
to the violation of the particle-hole symmetry. } and 
using its complex-conjugate $\tau_{\bf q}^*$ 
the time-dependence of the conjugate of (\ref{TD-Psi}) is given as 
\begin{equation}
\Psi_{\bf q}^*(t) = \Psi_{\bf q}^* e^{-t/\tau_{\bf q}^*}. 
\end{equation}
Since 
\begin{equation}
- {i \over 2} \lim_{t' \rightarrow t} 
\Big( {\partial \over \partial t} - {\partial \over \partial t'} \Big)
\Psi_{\bf q}^*(t) \Psi_{\bf q}(t') 
= - {\tau_{\bf q}^{''} \over (\tau_{\bf q}^{'})^2+(\tau_{\bf q}^{''})^2} 
\Psi_{\bf q}^*(t) \Psi_{\bf q}(t), 
\label{complex-tau} 
\end{equation}
we obtain\footnote{
The Drude-like formula (\ref{Drude-GL}) was extended 
to the case of the thermo-electric tensor by 
Howson, Salamon, Friedmann, Rice and Ginsberg: 
Phys. Rev. B {\bf 41}, 300 (1990) and 
Ausloos, Clippe and Patapis: 
Phys. Rev. B {\bf 46}, 5763 (1992). 
However, their discussion is incorrect 
so that they could not reach (\ref{alpha-xx--GL}). } 
within the linear order\footnote{
We have put $\tau_{\bf q}^{''} = 0$ 
except the numerator of the right-hand side of (\ref{complex-tau}). 
Thus the time-integral is identical to that leads to (\ref{Drude-GL}). } 
of $\gamma^{''} / \gamma^{'}$ 
\begin{align}
\tilde\alpha_{xx} 
& \fallingdotseq { e^* \over (m^*)^2 } {1 \over T} \sum_{\bf q} q_x^2 
\big\langle \Psi_{\bf q}^* \Psi_{\bf q} \big\rangle^2 
{\tau^{'}_{\bf q} \over 2} 
\times \Big( - {\gamma^{''} \over \gamma^{'}} \Big){1 \over \tau^{'}_{\bf q}} 
\nonumber \\ 
& = - 2 e^* T {\gamma^{''} \over \gamma^{'}} 
\sum_{\bf q} { \xi^4 q_x^2 \over (1 + \xi^2 q^2)^2 }, 
\label{alpha-xx--GL} 
\end{align}
where we have assumed $\gamma^{'} \gg |\gamma^{''}|$ 
with $\gamma \equiv \gamma^{'} + i \gamma^{''}$ 
and used (\ref{corr-length}). 

In 2D the ${\bf q}$-summation is performed as\footnote{
\begin{equation}
\int dx {x \over (x+b)^2} 
= \int dx {1 \over x+b} - b \int dx {1 \over (x+b)^2}. 
\nonumber 
\end{equation}
} 
\begin{equation}
\sum_{\bf q} { \xi^4 q_x^2 \over (1 + \xi^2 q^2)^2 } \fallingdotseq 
{1 \over 8 \pi d} \int_0^{q_c^2} dx {x \over (x + \xi^{-2})^2} 
\fallingdotseq {1 \over 8 \pi d} \ln(q_c\xi)^2, 
\label{log-sing} 
\end{equation}
so that\footnote{
Our result (\ref{alpha-2D}) is identical to 
(4.31) in \cite{new} and TABLE I in \cite{USH}. }
\begin{equation}
\alpha_{xx} \fallingdotseq 
{|e| \over 2 \pi d} {\gamma^{''} \over \gamma^{'}} 
\ln{T_\Lambda \over T-T_c}, 
\label{alpha-2D} 
\end{equation}
where we have used the Onsager relation (\ref{Onsager}) 
and introduced $T_\Lambda$ as $(q_c\xi)^2 \equiv T_\Lambda / (T-T_c)$. 

%%%%%%%%
\section{Cooper-Pair Transport: AL Process}
%%%%%%%%

%%%%%%%%%%%%%%%%%%%%%%%%%%%%%%%%%%%%%%%%%%
%\vskip 12mm
\begin{figure}[htbp]
\begin{center}
%\hskip -0.6cm
\includegraphics[width=10.0cm]{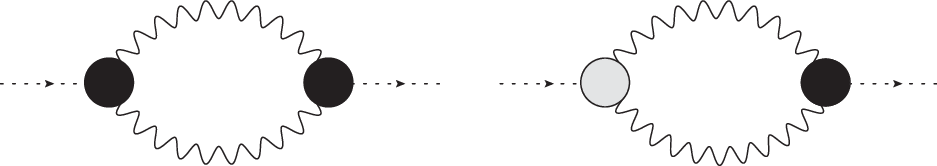}
\vskip 2mm
\caption{AL process of Cooper-pair transport. }
\label{fig:Fig-CP}
\end{center}
\end{figure}
%%%%%%%%%%%%%%%%%%%%%%%%%%%%%%%%%%%%%%%%%%

\noindent
Here we derive the GL results (\ref{sigma-2D}) and (\ref{alpha-2D}) 
from the AL process of the microscopic calculation. 
Such a process corresponds to the relaxation-time approximation 
neglecting the vertex correction. 
Since the effective interaction $L$ also obeys the same relations\footnote{
The Ward identities for $L$ are given as 
\begin{equation}
\sum_{\mu=0}^3 k_\mu \tilde\Delta_\mu^e(q,k) = 
2e L(q)^{-1} - 2e L(q+k)^{-1}, 
\nonumber 
\end{equation}
and 
\begin{equation}
\sum_{\mu=0}^3 k_\mu \tilde\Delta_\mu^Q(q+k,q) = 
q_0 L(q+k)^{-1} - (q_0+k_0) L(q)^{-1}, 
\nonumber 
\end{equation}
and we employ 
\begin{equation}
L({\bf q},i\omega_m\!=\!0)^{-1} = - N(0) 
\bigg[ \epsilon + \xi_0^2 \Big( {{\bf q}-{{\bf k}\over 2}} \Big)^2 \bigg], 
\nonumber 
\end{equation}
as (\ref{static-D}). } 
as (\ref{WIC-e}) and (\ref{WIC-Q}), 
we obtain the current vertices\footnote{
The frequency factor in (\ref{omega-rep}) 
appeared in (34) of [Uss]. 
However, such a factor cannot be derived 
from the preceding discussion in [Uss]. 
I shall discuss this point in the supplement noticed in the footnote 63. 

[Uss] $\equiv$ Ussishkin: Phys. Rev. B {\bf 68}, 024517 (2003). } for $L$ as 
\begin{equation}
\tilde\Delta_\mu^e = 4e N(0) \xi_0^2 q_\mu. 
\end{equation}
and 
\begin{equation}
\tilde\Delta_\mu^Q = 2 N(0) \xi_0^2 q_\mu 
\Big( i\omega_m + {i\omega_\lambda \over 2} \Big). 
\label{omega-rep} 
\end{equation}
These relations also hold 
even when we take the effect of impurity scattering into account.\footnote{
See, for example, \S 8.3 in \cite{new}. 
The value of $\xi_0$ becomes a function of the mean free path 
due to impurity scattering. } 

The charge response to electric field is determined by 
\begin{equation}
\Phi_{xx}^e({\bf k} \!=\! 0, i\omega_\lambda) = 
{1 \over \beta} \sum_{\bf q} \sum_m \tilde\Delta_x^e 
L(i\omega_m) L(i\omega_m+i\omega_\lambda) 
\tilde\Delta_x^e, 
\end{equation}
where we have used the simplified notation 
$L(i\omega_m) \equiv L({\bf q},i\omega_m)$. 
The discrete summation 
\begin{equation}
I^e(i\omega_\lambda) \equiv {1 \over \beta} \sum_m 
L(i\omega_m) L(i\omega_m+i\omega_\lambda), 
\label{sum-LL} 
\end{equation}
is transformed into the integral\footnote{
See, for example, \S 25 in [FW]. 
Since $\coth (z/2T) = 1 + 2 n(z)$, 
\begin{equation}
I^e(i\omega_\lambda) = \int_C {dz \over 4\pi i} \coth{z \over 2T} 
L(z)L(z+i\omega_\lambda). 
\nonumber 
\end{equation}
} 
\begin{equation}
I^e(i\omega_\lambda)
= \int_C {dz \over 2\pi i} n(z) 
L(z)L(z+i\omega_\lambda), 
\end{equation}
where the contour $C$ is shown in Fig.~5 and 
\begin{equation}
n(z) = {1 \over e^{\beta z} - 1}, 
\end{equation}
is the Bose distribution function. 
Employing 
\begin{equation}
L^R(x) - L^A(x) = 2 i \, {\rm Im}L^R(x), 
\end{equation}
and the relation $ n(z - i\omega_\lambda) = n(z) $
led by $ e^{-i\beta\omega_\lambda} = 1 $ 
we obtain 
\begin{equation}
I^e(\omega+i\delta) 
= \int_{-\infty}^\infty {dx \over \pi} n(x) 
\Big[ L^R(x + \omega) + L^A(x - \omega) \Big] 
{\rm Im}L^R(x), 
\end{equation}
by repeating the procedure leading to (\ref{Int-e-final}). 
The contribution linear in $\omega$ is obtained as (\ref{I-e-omega})$:$ 
\begin{equation}
I^e(\omega+i\delta) - I^e(i\delta) \fallingdotseq 
i\omega \int_{-\infty}^\infty {dx \over \pi} 
\Big( - {\partial n(x) \over \partial x} \Big) \Big[ {\rm Im} L^R(x) \Big]^2, 
\end{equation}
where 
\begin{equation}
{\rm Im}L^R(x) = - {1 \over N(0)} 
{\tau_0 x \over \eta^2 + (\tau_0 x)^2}, 
\end{equation}
with 
$\eta \equiv \epsilon + \xi_0^2 {\bf q}^2$. 
Thus the resulting DC conductivity is 
\begin{align}
\sigma_{xx} 
& = \sum_{\bf q} \big( \tilde\Delta_x^e \big)^2 
\int_{-\infty}^\infty {dx \over \pi} 
\Big( - {\partial n(x) \over \partial x} \Big) \Big[ {\rm Im} L^R(x) \Big]^2 
\nonumber \\ 
& = {1 \over 4 \pi T} \sum_{\bf q} \big( \tilde\Delta_x^e \big)^2 
\int_{-\infty}^\infty dx 
\Big[ \sinh{x \over 2T} \Big]^{-2} \Big[ {\rm Im} L^R(x) \Big]^2. 
\end{align}
In order to discuss the effect of the low-energy critical fluctuation 
at finite temperature 
it is enough to use the high-temperature expansion\footnote{
The high-temperature expansion is compatible with the Kubo formula 
(\ref{sigma-Kubo}) and (\ref{alpha-Kubo}) for classical variables. }
$\sinh(x/2T) \fallingdotseq x/2T$ 
under the assumption $\epsilon \ll 1$. 
By performing the integral\footnote{
Since 
\begin{equation}
\Big[ \sinh{x \over 2T} \Big]^{-2} \Big[ {\rm Im} L^R(x) \Big]^2 
\fallingdotseq 
\Big( {2T \over N(0)} \Big)^2 
\Big( {\tau_0 \over \eta^2 + (\tau_0 x)^2} \Big)^2, 
\nonumber 
\end{equation}
the integral is evaluated as 
\begin{equation}
\int_{-\infty}^\infty dx 
\Big[ \sinh{x \over 2T} \Big]^{-2} \Big[ {\rm Im} L^R(x) \Big]^2 
\fallingdotseq 
\Big( {2T \over N(0)} \Big)^2 2 \tau_0 
\int_{0}^\infty {dy \over (y^2 + \eta^2)^2}. 
\nonumber 
\end{equation}
Using the definite integral appeared in the footnote for (\ref{sigma-RTA}) 
we obtain 
\begin{equation}
\int_{0}^\infty {dy \over (y^2 + \eta^2)^2} 
= {\pi \over 4} {1 \over \eta^3}. 
\nonumber 
\end{equation}
} 
we obtain 
\begin{equation}
\sigma_{xx} = 8 e^2 \tau_0 T \sum_{\bf q} 
{ \xi_0^4 q_x^2 \over (\epsilon + \xi_0^2 {\bf q}^2)^3 }. 
\end{equation}

In 2D the ${\bf q}$-summation is performed as\footnote{
By the same procedure as the footnote for (\ref{int-AL}) 
\begin{equation}
\int_{0}^\infty dx {x \over (x + \epsilon)^3} 
= {1 \over 2\epsilon}. 
\nonumber 
\end{equation}
} 
\begin{equation}
\sum_{\bf q} { \xi_0^4 q_x^2 \over (\epsilon + \xi_0^2 {\bf q}^2)^3 } 
= {1 \over 8 \pi d} \int_0^\infty dx {x \over (x + \epsilon)^3} 
= {1 \over 16 \pi d} {1 \over \epsilon}, 
\end{equation}
so that\footnote{
Here we have used $\tau_0 = \pi / 8T$ 
derived within the ladder approximation. 
See, for example, \S 6.2 in \cite{new}. } 
\begin{equation}
\sigma_{xx} 
= {e^2 \over 2 \pi d} {\tau_0 T \over \epsilon} 
= {e^2 \over 16 d} {1 \over \epsilon}. 
\end{equation}

The heat response to electric field is determined by 
\begin{equation}
\Phi_{xx}^Q({\bf k} \!=\! 0, i\omega_\lambda) = 
{1 \over \beta} \sum_{\bf q} \sum_m \tilde\Delta_x^Q 
L(i\omega_m) L(i\omega_m+i\omega_\lambda) 
\tilde\Delta_x^e. 
\label{heat-response} 
\end{equation}
The discrete summation 
\begin{equation}
I^Q(i\omega_\lambda) \equiv {1 \over \beta} \sum_m 
\Big( i\omega_m + {i\omega_\lambda \over 2} \Big) 
L(i\omega_m) L(i\omega_m+i\omega_\lambda), 
\end{equation}
is transformed into the integral 
\begin{equation}
I^Q(\omega+i\delta) 
= \int_{-\infty}^\infty {dx \over \pi} n(x) 
\Big[ \Big( x + {\omega \over 2} \Big) 
L^R(x + \omega) 
+ \Big( x - {\omega \over 2} \Big) 
L^A(x - \omega) \Big] 
{\rm Im}L^R(x). 
\end{equation}
In the same manner as (\ref{Int-Q-linear}) we obtain 
\begin{equation}
I^Q(\omega+i\delta) - I^Q(i\delta) \fallingdotseq 
i\omega \int_{-\infty}^\infty {dx \over \pi} 
\Big( - {\partial n(x) \over \partial x} \Big) 
x \Big[ {\rm Im}L^R(x) \Big]^2. 
\label{Int-Q-linear-CP} 
\end{equation}
Employing the high-temperature expansion $n(x) \fallingdotseq T/x$, 
(\ref{Int-Q-linear-CP}) reduces to  
\begin{equation}
I^Q(\omega+i\delta) - I^Q(i\delta) \fallingdotseq 
{\omega T \over 4 \pi} 
\int_{-\infty}^\infty dx 
\Big[ L^R(x) - L^A(x) \Big] 
{ L^R(x) - L^A(x) \over i x }. 
\label{Int-Q-highT} 
\end{equation}
Here 
\begin{equation}
L^R(x) = - {1 \over N(0)} 
{1 \over \eta - i \tau_0 x}, 
\ \ \ \ \ \ \ \ \ \ 
L^A(x) = - {1 \over N(0)} 
{1 \over \eta + i \tau_0^* x}, 
\end{equation}
with $\tau_0 \equiv \tau_1 + i \tau_2$ and 
$\tau_0^* \equiv \tau_1 - i \tau_2$ 
and we can put $\tau_1 \gg |\tau_2|$. 
The right-hand side of (\ref{Int-Q-highT}) is evaluated exactly 
by the residue as\footnote{
The integral 
\begin{equation}
I \equiv 
\int_{-\infty}^\infty dx 
\Big[ L^R(x) - L^A(x) \Big] 
{ L^R(x) - L^A(x) \over i x }, 
\nonumber 
\end{equation}
is written as 
\begin{equation}
I = {\tau_0^* + \tau_0 \over N(0)^2} 
\int_{-\infty}^\infty dx 
\Big[ {1 \over \eta - i \tau_0 x} - {1 \over \eta + i \tau_0^* x} \Big] 
{1 \over (\eta - i \tau_0 x)(\eta + i \tau_0^* x)}, 
\nonumber 
\end{equation}
and is evaluated by the residue as 
\begin{equation}
I = {2\pi \over N(0)^2} {\tau_0^* - \tau_0 \over \tau_0^* + \tau_0} 
{1 \over \eta^2}, 
\nonumber 
\end{equation}
where the position of the pole is determined by the condition 
$\tau_1 \gg |\tau_2|$. }
\begin{equation}
- i \omega {T \over 2 N(0)^2} 
{\tau_2 \over \tau_1} {1 \over \eta^2}. 
\label{I_1=0} 
\end{equation}
Thus the analytic continuation of (\ref{heat-response}) is given as\footnote{
Since $\tau_2 < 0$ in 3D for our free electron dispersion (\ref{xi(p)}), 
(\ref{thermo-power}) implies that 
$\tilde\alpha_{xx}$ is proportional to the charge of Cooper pairs. 
For example, (2.27) in 
Fukuyama, Ebisawa and Tsuzuki: Prog. Thoer. Phys. {\bf 46}, 1028 (1971) 
shows $\tau_2 < 0$ 
after the correction pointed out by several authors: 
its right-hand side should be multiplied by $-1$. } 
\begin{equation}
\Phi_{xx}^Q({\bf k} \!=\! 0, \omega+i\delta) = (-i\omega) 
4eT {\tau_2 \over \tau_1} 
\sum_{\bf q} { \xi_0^4 q_x^2 \over (\epsilon + \xi_0^2 {\bf q}^2)^2 }. 
\label{thermo-power} 
\end{equation}

In 2D the ${\bf q}$-summation is performed as (\ref{log-sing})$:$ 
\begin{equation}
\sum_{\bf q} { \xi_0^4 q_x^2 \over (\epsilon + \xi_0^2 {\bf q}^2)^2 } 
\fallingdotseq 
{1 \over 8 \pi d} \int_0^{x_c} dx {x \over (x + \epsilon)^2} 
\fallingdotseq {1 \over 8 \pi d} \ln{x_c \over \epsilon}, 
\end{equation}
so that 
\begin{equation}
\tilde\alpha_{xx} \fallingdotseq 
- {e \over 2 \pi d} T {\tau_2 \over \tau_1} \ln{x_c \over \epsilon}. 
\end{equation}
By the Onsager relation (\ref{Onsager}) we finally obtain 
\begin{equation}
\alpha_{xx} = {1 \over T} \tilde\alpha_{xx} \fallingdotseq 
{|e| \over 2 \pi d} {\tau_2 \over \tau_1} 
\ln{T_\Lambda \over T-T_c}, 
\end{equation}
which is equivalent to (\ref{alpha-2D}) 
where 
$\tau_2 / \tau_1 = \gamma^{''} / \gamma^{'}$ 
and $x_c$ is chosen as $x_c/\epsilon \equiv T_\Lambda / (T-T_c)$. 

%%%%%%%%
\section{Ward Identities: Ladder Approximation}
%%%%%%%%

It is instructive to show 
how the Ward identity is satisfied in the ladder approximation 
that is usually employed 
to obtain the explicit form of the Cooper-pair propagator. 
In particular it is very important to clarify 
how the Ward identity for heat current vertex is satisfied, 
because it will clear away some confusions\footnote{
For example, 
if we employ the heat current vertex for electrons 
with the frequency factor $i\varepsilon_n + i\omega_\lambda / 2$ 
in perturbation expansion, we obtain some wrong results. 
The frequency factor should be accompanied by the full propagator. 
It should not be accompanied by the free propagator. } 
seen in the literatures. 
Such a clarification becomes inevitably complicated 
so that it will be uploaded as a separate supplementary note.\footnote{
I shall upload a note entitled 
{\it A Diagrammer's Note on Superconducting Fluctuation Transport 
for Beginners: Supplement} besides the three notes in series. } 

%%%%%%%%
\section{Current Vertices: Perturbational Analysis}
%%%%%%%%

As has been shown in this note 
we do not have to calculate the current vertices for Cooper pairs 
to obtain $\sigma_{xx}$ and $\alpha_{xx}$. 
However, it is instructive to construct the current vertices 
by perturbational calculation.\footnote{
If we perform the perturbational calculation of the current vertex 
without the ground as the Ward identity, 
it is difficult to convince ourselves of the validness of the result. 
For example, there are many errors in the calculation of vertices by 
Varlamov and Livanov: Sov. Phys. JETP {\bf 71}, 325 (1990) and 
{\bf 72}, 1016 (1991), 
besides the fatal error, 
concerning the $\lq\lq$cancelation", corrected by [RS]. 

[RS] $\equiv$ Reizer and Sergeev: Phys. Rev. B {\bf 50}, 9344 (1994). } 
Such a construction for the heat current vertex is complicated\footnote{
The factor of the heat current vertex (45) in [RS] 
is incorrect but this result is supported 
by Serbyn, Skvortsov, Varlamov and Galitski: 
Phys. Rev. Lett. {\bf 102}, 067001 (2009). 
I do not understand the reason of the support. 
On the other hand, the heat current vertex (21) in [Uss], 
which is cited as (10.23) in \cite{old} 
and corresponds to (\ref{omega-rep}) with $\omega_\lambda = 0$, 
is insufficient to obtain $\alpha_{xx}$. } 
so that it will be included in the supplementary note 
noticed in the previous section. 

%%%%%%%%
\section{Exercise}
%%%%%%%%

Find and correct the errors in this note. 
(Do not trust the results in this note before you check them.)

\vskip 8mm

\noindent
{\bf{\Large Acknowledgements}}

\vskip 3mm
\noindent
I am grateful to Professors Hidetoshi Fukuyama and Kazumasa Miyake. 
I became a diagrammer trained by Hidetoshi Fukuyama 
and this note was prepared to answer the questions 
raised by Kazumasa Miyake. 
I have been benefitted much from {\boldmath$\mathsf{arXiv}$} 
so that I will be pleased 
if this note can increase the value of {\boldmath$\mathsf{arXiv}$}. 
This note is humble but intended to celebrate 
the 100-th anniversary of the discovery of superconductivity.  

%----------------------------------------------------------------

%%%%%%%%
\section{Remarks on Heat Current}
%%%%%%%%

In the footnote 16 we have assumed that 
${\cal L_{\rm int}}$ does not contain the derivative of $\psi_\sigma^\dag$ and $\psi_\sigma$. 
Since we are interested in the simple case of contact interaction, we are safe to adopt such an assumption. 

However, if we consider an interaction with finite range, 
some heat currents not described by the equation (63) arise. 
See Appendix B in the paper: Ogata, Fukuyama, J. Phys. Soc. Jpn. 88 (2019) 074703 
for a detailed discussion. 
This correction to (63) is due to the derivative term in ${\cal L_{\rm int}}$. 
Such a correction was cautioned in the preprint: 
Moreno, Coleman, arXiv:cond-mat/9603079v2. 
This type of heat current already appeared as the equation (26) 
in the paper: Jonson, Mahan, Phys. Rev. B 42 (1990) 9350. 

You might feel the gap between the equations (86) and (88). 
Here we should remember the derivative in terms of $\tau$ as defined in the equation (8) 
where the $\tau$-dependence is determined by $K = H - \mu N$. 
Thus (88) incorporates the contribution of the chemical potential besides the energy current. 

The Ward identity for heat-current vertex is discussed in arXiv:1309.4257v2. 

%%%%%%%%
\section{Gauge Theory of Heat}
%%%%%%%%

The charge-conservation law is given in (62). 
This conservation law is deeply related to the gauge invariance. 
Schrieffer [{\it Theory of Superconductivity} (Benjamin/Cummings, 1964) \S 8-5] 
discussed the relation in terms of the electromagnetic response. 
The relation is directly demonstrated in terms of the action. 
The action for the interaction between electrons and photons is given by 
$$
S = \int {\rm d}^4 x J^\mu A_\mu .
$$
Under the gauge transformation $A_\mu \rightarrow A_\mu + \partial_\mu \chi$ 
the action is deformed as 
$$
\delta S = \int {\rm d}^4 x J^\mu \partial_\mu \chi = - \int {\rm d}^4 x \partial_\mu J^\mu \chi
$$
where we have exploited the integration by parts 
and assumed the vanishing contribution from the surface at infinity. 
The gauge invariance of the action ($\delta S = 0$ for any $\chi$) 
results in the charge-conservation law ($\partial_\mu J^\mu = 0$). 

The heat-conservation law (66) has the same form\footnote{
The charge conservation (62) and the energy conservation (64) lead to the heat conservation (66).
} as (62). 
Thus there exists a gauge theory of heat. 
We can introduce the gauge field\footnote{
See Saleem, Schwingenschl\"ogl, Manchon, Phys. Rev. B 109 (2024) 134415 for a review.} 
(the scalar potential $\phi$ and the vector potential $\vec A$) for heat. 
The thermal electric field $\vec E$ and the thermal magnetic field $\vec B$ 
are given by ${\vec E} = -\nabla\phi - {\dot {\vec A}}$ and ${\vec B} = \nabla \times {\vec A}$ 
as in the case of usual electromagnetic gauge theory. 
Here $\vec E$ is related to the temperature gradient as ${\vec E} = - \nabla T / T$. 
\color{red}
In this framework, the heat current $\vec J$ is expressed 
as ${\vec J} =  {\vec J}_{\vec E} + {\vec J}_{\vec B}$ 
where ${\vec J}_{\vec E}$ is related to $\vec E$ and ${\vec J}_{\vec B}$ is related to $\vec B$. 
${\vec J}_{\vec E}$ is driven by the temperature gradient. 
${\vec J}_{\vec B}$ is expressed as 
$$
{\vec J}_{\vec B} = \nabla \times {\vec M}
$$
and exists even in the equilibrium state without the temperature gradient. 
Here $\vec M$ is the heat magnetization 
given by ${\vec M} = - \partial \Omega / \partial {\vec B}$ from the free energy $\Omega$. 
The heat-magnetization current ${\vec J}_{\vec B}$ is not captured by ordinary Kubo formula. 
\color{black}

%----------------------------------------------------------------

\begin{thebibliography}{99}
\bibitem{xxx} 
In the following 
I only list the references that you must read. 
Neither originality nor priority is considered here. 
Other references are cited in the footnotes.
\end{thebibliography}
\end{document}